\pdfoutput = 1

\documentclass[journal]{IEEEtran}
\ifCLASSINFOpdf
\else
\fi
\usepackage{amsmath}
\usepackage{algorithmic}

\usepackage{array}
\usepackage{url}

\usepackage{graphicx}
\usepackage{gensymb}
\usepackage[normalem]{ulem}
\useunder{\uline}{\ul}{}
\usepackage{booktabs}
\usepackage{multirow}
\usepackage{siunitx}
\usepackage{tabulary}
\usepackage{booktabs}
\usepackage{xcolor}
\usepackage{graphicx}
\usepackage{amsmath}
\usepackage{threeparttable}
\usepackage{enumitem}
\usepackage{longtable}
\usepackage{xcolor}
\usepackage{graphicx}
\usepackage{amsmath}
\usepackage{threeparttable}
\usepackage{enumitem}
\usepackage{listings}
\usepackage{subfigure}
\usepackage{todonotes}
\usepackage[T1]{fontenc}
\usepackage{svg}
\usepackage[normalem]{ulem} 
\usepackage{graphics} 

\begin{document}
%
\title{Vega: A 10-Core SoC for IoT End-Nodes with DNN Acceleration and Cognitive Wake-Up From MRAM-Based State-Retentive Sleep Mode}
%
%

\author{Davide~Rossi,~\IEEEmembership{Member,~IEEE,}
        Francesco~Conti,~\IEEEmembership{Member,~IEEE,}
        Manuel~Eggimann,~\IEEEmembership{Member,~IEEE,}
        Alfio~Di~Mauro,~\IEEEmembership{Member,~IEEE,}
        Giuseppe~Tagliavini,~\IEEEmembership{Member,~IEEE,}
        Stefan~Mach,
        Marco~Guermandi,
        Antonio~Pullini,
        Igor~Loi,
        Jie~Chen,
        Eric~Flamand,
        Luca~Benini,~\IEEEmembership{Fellow,~IEEE.}

\thanks{D. Rossi, F. Conti, G. Tagliavini, M.Guermandi, J. Chen, L. Benini are with University of Bologna, Italy. M. Eggiman, A. Di Mauro, S. Mach, L. Benini are with ETH Zurich, Switzerland. A. Pullini, I. Loi, J. Chen, E. Flamand are with GreenWaves Technologies, Grenoble, France.}
\thanks{This work was supported in part by the EU Horizon 2020 Research and Innovation projects OPRECOMP (g.a. no. 732631) and WiPLASH (g.a. no. 863337) and by the ECSEL Horizon 2020 project AI4DI (g.a. no. 826060).}
}

\maketitle

\begin{abstract}

The Internet-of-Things requires end-nodes with ultra-low-power always-on capability for a long battery lifetime, as well as high performance, energy efficiency, and extreme flexibility to deal with complex and fast-evolving near-sensor analytics algorithms (NSAAs). We present Vega, an IoT end-node SoC capable of scaling from a 1.7 $\mathrm{\mu}$W fully retentive cognitive sleep mode up to 32.2 GOPS (@ 49.4 mW) peak performance on NSAAs, including mobile DNN inference, exploiting 1.6 MB of state-retentive SRAM, and 4 MB of non-volatile MRAM. To meet the performance and flexibility requirements of NSAAs, the SoC features 10 RISC-V cores: one core for SoC and IO management and a 9-cores cluster supporting multi-precision SIMD integer and floating-point computation. Vega achieves SoA-leading efficiency of 615 GOPS/W on 8-bit INT computation (boosted to 1.3TOPS/W for 8-bit DNN inference with hardware acceleration). On floating-point (FP) compuation, it achieves SoA-leading efficiency of 79 and 129 GFLOPS/W on 32- and 16-bit FP, respectively. Two programmable machine-learning (ML) accelerators boost energy efficiency in cognitive sleep and active states, respectively.

\end{abstract}


%
\IEEEpeerreviewmaketitle

\section{Introduction}

An increasing amount of near-sensor data analytics applications require inexpensive battery-operated micro-systems able to sense the environment and transmit meaningful, highly semantic compressed data to the cloud wirelessly. The tight constraints in terms of low-power in sleep mode, coupled with extreme performance and energy efficiency in active mode, calls for a new class of ultra-low-power microcontrollers (MCUs), namely, IoT processors. These devices require a large state-retentive memory to autonomously wake up when always-on ultra-low-power sensors detect a specific condition. Following wake-up events, more capable sensors and computing units can be activated to perform fully programmable complex near-sensor analytics, including modern Deep Neural Networks (DNNs) models. This approach enables the extraction of meaningful information from the sensor data locally before transmitting it, avoiding data deluge in the cloud.


Recent research on ULP MCUs design focused on main building blocks, such as memories \cite{HAINE_SRAM_ESSCIRC_2017}, standard cells \cite{REYNDERS_TCAS_II_2012} and embedded power management \cite{MONO_JSSC_2020}. On the other hand, while more traditional low-end IoT end-nodes are targeted to low-bandwidth sensors (e.g., temperature and pressure) and require limited compute capabilities, an increasing number of applications rely on embedding much more intelligence at the edge. Dedicated solutions explored in the last few years deliver high performance and efficiency, mainly focusing on inference \cite{ENVISION_ISSCC_2017, UNPU_ISSCC_2018}, and training \cite{GANPU_JSSC_2021} of DNNs, exploiting low-precision and tunable-precision arithmetic to adapt to the requirements of applications while minimizing energy consumption \cite{IBM_ISSCC_2021}. Although the efficiency of dedicated hardware accelerators is orders of magnitude larger than that of previously mentioned MCUs, the large variety and fast evolution of near-sensor data analytics algorithms running on IoT end-nodes cannot be satisfied by specialized and inflexible accelerators.


In this context, this work provides a significant step forward in high-performance, Parallel Ultra-Low-Power (PULP) IoT processors, presenting Vega. Vega introduces key contributions in two areas: always-on cognitive operation augmented by non-volatile memory support and highly dynamic digital signal processing. First, the proposed SoC features 4 MB of non-volatile MRAM coupled with a fully programmable cognitive wake-up unit based on the Hyper Dimensional Computing (HDC) paradigm \cite{KANERVA_CC_2009}. This non-volatile cognitive wake-up architecture enables the probing of ultra-low-power sensor data with power consumption as low as 1.7 $\mu$W and wakes up the system from a full memory retentive state. Second, Vega enables the highly dynamic exploitation of multiple data formats, from a few bit-width integer to full precision floating-point (FP). Thus, application developers can seamlessly tune the precision and dynamic range of portions of algorithms, matching them with the rich set of data formats natively available on the hardware. We demonstrate the capabilities of the Vega SoC on a wide range of Near-Sensor Analytic Applications (NSAA) in the bio-signal, audio/vibration, and imaging domains as well as inference of DNN, showing significant improvement in flexibility, performance, and efficiency over the state of the art.

\begin{figure*}[t!]
     \centering
     \includegraphics[width=\linewidth]{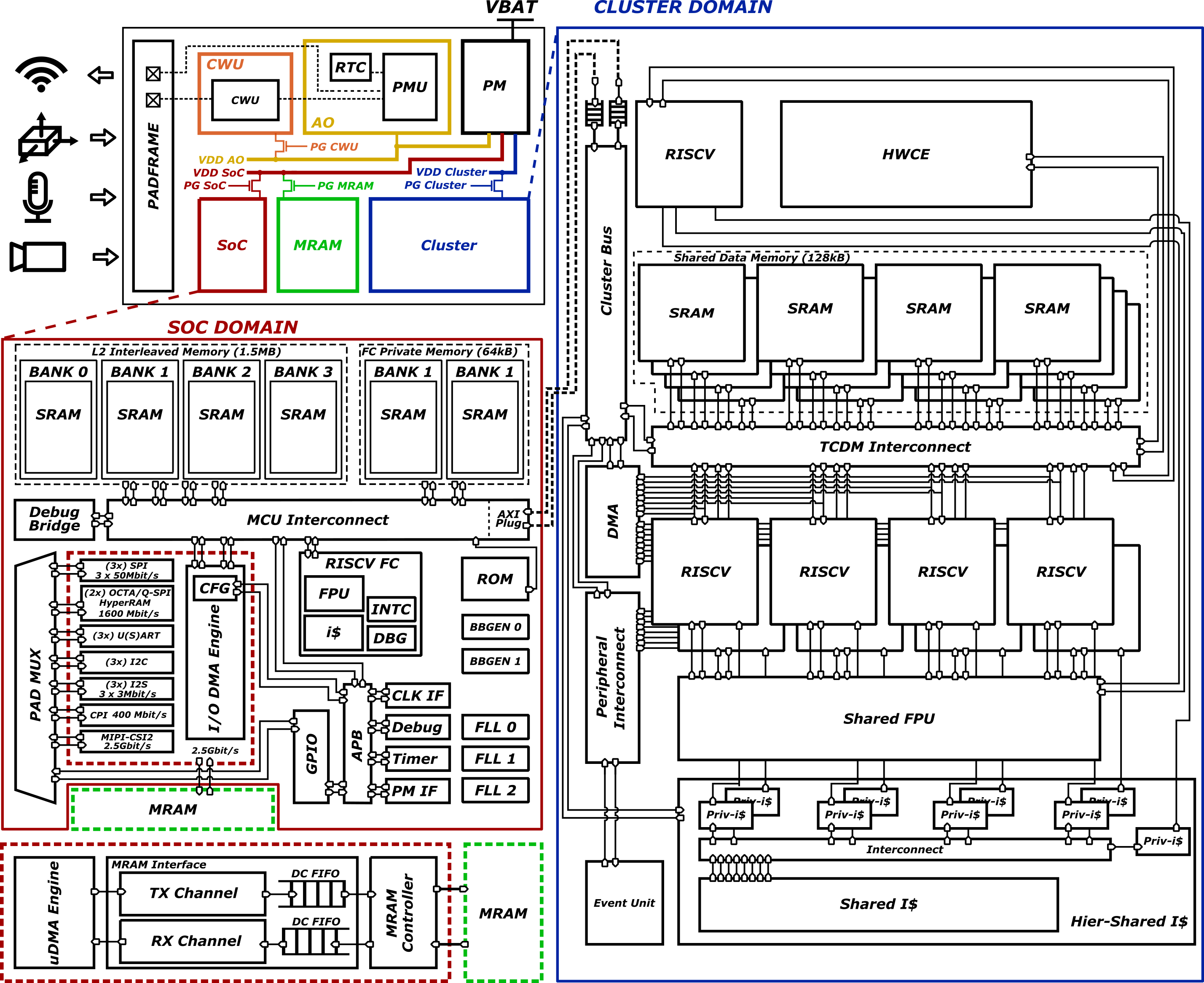}
     \caption{Vega SoC Architecture}
     \label{fig:archi}
     \vspace{-5mm}
\end{figure*}

\section{VEGA SoC Architecture}
\label{sec:archi}

As shown in Fig.~\ref{fig:archi}, Vega consists of four main switchable power domains: the SoC domain includes a single-core MCU served by 1.7 MB of SRAM memory and several peripherals, the cluster domain includes the programmable parallel accelerator, a domain including 4 MB of non-volatile MRAM, and a fourth domain including the Cognitive Wake-Up (CWU) unit. 

\subsection{Non-Volatile SoC Subsystem}

The SoC Domain is an advanced MCU featuring a RISCV processor named Fabric Controller (FC), served by 1.7 MB of embedded SRAM and several peripherals described in Fig. \ref{fig:archi}, including a 1.6 Gbit/s HyperBus/OCTA SPI/SDIO Double Data Rate (DDR) interface supporting external DRAM and Flash memories. For instance, the interface supports Cypress Semiconductor's HyperRAM and HyperFlash memories \cite{HyperRAM}, AP Mmemory IoT RAMs \cite{APMemory}, and external Secure Digital Input Output (SDIO) cards.

To allow peripherals to transfer data independently from FC, Vega implements an I/O subsystem in which each peripheral has a dedicated DMA channel enabling direct, autonomous data transfer to/from the L2 memory \cite{WOLF_JSSC_2019}.
A 4 MB non-volatile Magnetoresistive Random Access Memory (MRAM) resides in an independent switchable power domain. The MRAM is connected to the L2 memory through an auxiliary IO DMA channel and managed just like a peripheral. A dedicated controller manages the specific protocol conversion operations required to access the MRAM in write and read mode, completely abstracting to the end-user the complexity of the specific protocol. During the read operation, the MRAM can operate at a frequency up to 40 MHz, delivering bandwidth of 2.5 Gbit/s through its 78-bit interface (including 14-bit ECC). The two main applications of MRAM within Vega SoC are the storage of read-only parameters of machine learning applications and program code. Except for the I/O DMA, the master resources of the SoC can only access MRAM data after being moved to the L2 memory.

\begin{figure*}[t!]
  \centering
  \includegraphics[width=\linewidth]{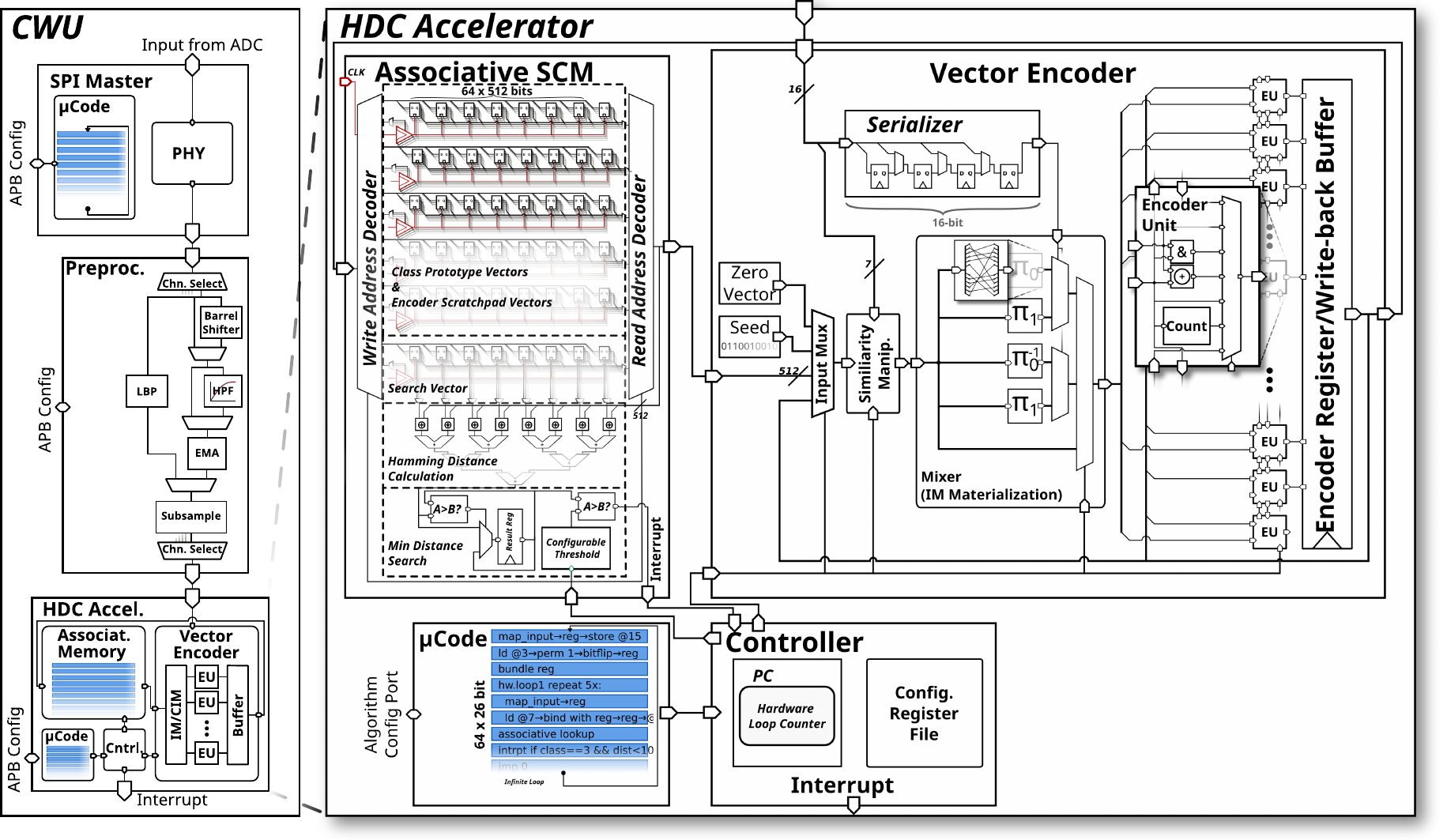}
  \caption{Architectural overview of the Cognitive Wake-up Unit consisting of autonomous SPI master module, configurable preprocessor and HDC Accelerator.}
  \label{fig:cwu_overview}
  \vspace{-5mm}
\end{figure*}

The L2 memory memory consists of 4 word-level interleaved banks for a total of 1.5MB, plus 64 kB of private memory for the core. 
This architecture provides an overall bandwidth of 6.7 GBytes/s to the peripherals and accelerators in the system.
To retain the SoC program and data in sleeping mode, the physical SRAM banks can selectively be configured in retentive mode, leading to retention power ranging from 1.2 to 112 $\mu$W for 16 kB to 1.6 MB of state-retentive L2 SRAM. Hence, once the SoC is woken-up from sleep mode, a \textit{warm boot} can either be performed from L2 SRAM or from the MRAM. In the former case, some power consumption is required for preserving state-retention of SRAMs; in the latter case, sleep power for data retention is zero, but the program must be restored into L2 after wake-up. Hence, depending on the duty cycle and wake-up latency requirement of the target IoT application, one or the other approach can be selected.

\subsection{Cognitive Wake-Up Unit}

While Vega provides highly energy-efficient compute performance in active mode using its programmable cluster, the TinyML \cite{Cho2019, Shan2020, Zhao2020} power envelope for self-sustainable always-on signal processing applications cannot be met in full active mode.
These applications require aggressive duty cycling and intelligent wake-up logic to detect events of interest.
However, the threshold-based wake-up systems used by most applications do not provide a small enough false-positive rate at an acceptable false-negative rate for effective power saving \cite{ROVERE_JETCAS_2018}.
To deal with those kinds of applications, Vega contains a programmable cognitive wake-up unit (CWU) that performs end-to-end machine learning on external sensor data and triggers the embedded power management unit (PMU) to power up the cluster for more advanced data analytics.
The CWU is designed to operate entirely autonomously: after its initial configuration, the CWU continuously processes and classifies external sensor data without any further interaction of the cores.

Figure \ref{fig:cwu_overview} gives a hierarchical overview of the CWU.
It consists of three main components:
\begin{itemize}
\item Programmable SPI master peripheral
\item Low-power preprocessor module
\item Configurable HDC accelerator
\end{itemize}

The dedicated SPI master peripheral acts as the IO interface to interact with external sensors.
It supports all four SPI phase and polarity configurations and controls up to four chip select signals.
Complex transaction patterns involving wait cycles and arbitrary read and write transactions with multiple external devices can be programmed utilizing an integrated micro-instruction memory that executes the configured access pattern in an endless loop.
The preprocessor module optionally performs lightweight data preprocessing and data format conversion of the digital sensor data on up to eight independent channels.
It supports data width conversion, offset removal, low-pass filtering, subsampling, and local-binary-pattern (LBP) filtering \cite{LBP}.
The offset removal and low-pass filters are based on an exponential moving average filter with a configurable decay rate to save area and power.

The core of the CWU is \textit{Hypnos}, a programmable hardware accelerator for HDC. HDC is a brain-inspired computing paradigm for machine learning that operates on high-dimensional holistic representations of the input data~\cite{KANERVA_CC_2009}. HDC has been proven to achieve competitive accuracy performance in various domains like biosignals processing~\cite{MOIN_NATURE_EL_2021}, DNA sequencing~\cite{IMANI_BHI_2018}, language classification~\cite{JOSHI_QI_2017}, and vehicle classification~\cite{KLEYKO_ICCOINS_2014}.
With its few-shot learning capability~\cite{RAHIMI_MONA_2017} and inherent error-resiliency in the presence of random bit flips~\cite{IMANI_HPCA_2017} HDC is an ideal candidate for an online-trainable wakeup circuit operating at low voltage.

In the context of \textit{Hypnos}, HDC is used to encode a time-series from one or several digital sensor channels to a high-dimensional binary \textit{search vector} using a small set of bit-wise and thus well parallelizable operations.
Then, this search vector is compared with so-called \textit{prototype vectors} representing the individual classes of interest in associative memory (AM).

In Vega, \textit{Hypnos} operates on 512, 1024, 1536, or 2048-bit HD vectors with a 512-bit wide datapath. The Vector Encoder module is responsible for encoding low-dimensional input data to high-dimensional vectors (HD-vectors).
It performs so-called \emph{item memory} (IM) mapping and the operation primitives of HDC like bundling and binding in an iterative manner~\cite{KANERVA_CC_2009}. Instead of employing a ROM-based IM that stores the mapping from low-dimensional input to pseudo-random HD-vectors, \textit{Hypnos} uses IM "rematerialization" by using a set of four hardwired random permutations. Item memory mapping is thus performed by iteratively applying random permutations on a hardwired pseudo-random seed vector with the bits from the serialized input word acting as select signals to switch between the different permutations. In this manner, \textit{Hypnos} can materialize an IM HD-vector in $D$ cycles, where $D$ denotes the configurable input data width from the pre-processor. While IM maps values from the low-dimensional input space to quasi-orthogonal HD-vectors, continuous item memory mapping (CIM) encodes input values so that low euclidean distance in the input space is mapped to low hamming distance in HD space~\cite{RAHIMI_JPROC_2019}. In most HDC algorithms, IM mapping is used to encode channel labels and CIM to encode the channel values to preserve the similarity of the values. To support CIM mapping, the vector encoder contains the similarity manipulator module that allows flipping of a configurable number of bits from an input HD-vector.

The bitwise HDC operations are realized in the Encoder Units (EU), of which \textit{Hypnos} contains 512 instances (one for each bit).
Each EU contains the logic for XOR/AND/NOT operation and a saturating bidirectional 8-bit counter that counts the number of encountered ones and zeros for bundling. A 32 kbit standard-cell based associative memory (AM) can hold up to 16 HD-vectors and acts as both a scratchpad memory to store intermediate HD-vectors from the Vector Encoder and store the final prototype- and search-vector during the associative lookup operation. The vector encoder can fetch HD-vectors from the AM and stores the result of each encoding round in a 512-bit wide register that can be fed back as the source for the next encoding cycle or written back to the AM.

The AM uses latches as storage primitives with a single integrated clock gate (ICG) per ROW acting as a write-enable line. For associative lookup (search for entry with minimal Hamming distance to the search word), the AM sequentially compares each memory row with the search vector and combinationally calculates the Hamming distance. This operation is usually invoked last in most HDC based classification algorithms to compare an encoded search vector against a set of prototype vectors that represent the individual classes of interest~\cite{RAHIMI_JPROC_2019}. Within our AM architecture, the index and Hamming distance of the most similar item are compared against a configurable threshold and target index. If the result of the lookup is of sufficient similarity to the target index, an interrupt is raised to trigger the wake-up sequence within Vega's PMU.

Since the optimal HDC encoding algorithm, that is, the sequence of encoding operations within the Vector Encoder, is application-dependent, the CWU contains another 64\texttimes26-bit SCM to encode the HDC algorithm in a sequence of compact micro-code instructions. The lightweight controller fetches these instructions in an infinite loop and reconfigures AM and Vector Encoder accordingly in each cycle. Thus, the CWU can be configured to run classification algorithms on the preprocessed sensor data and trigger wake-up sequences in Vega's PMU in a fully autonomous manner. The design is intended to operate at very low frequencies in the order of dozens of kHz. Thus the module was realized in UHVT logic to minimize leakage power. The design occupies a total area of 0.147 $mm^2$ and operates at 0.6V.

\begin{table}
  \centering
  \caption{Implementation details and power consumption of the Cognitive Wakeup Unit}
  \begin{tabular}[h]{lcc}
    \toprule
    & $f_{clk}=\SI{32}{\kilo\hertz}$ & $f_{clk}=\SI{200}{\kilo\hertz}$ \\
    \midrule
    Max. Samp. Rate & \SI{150}{SPS/Channel} & \SI{1}{\kilo SPS/Channel} \\
    $P_{\textrm{dynamic, datapath}}$ & \SI{0.99}{\micro \watt} & \SI{6.21}{\micro \watt} \\
    $P_{\textrm{dynamic, SPI pads}}$ & \SI{1.28}{\micro \watt} & \SI{8.00}{\micro \watt} \\
    $P_{\textrm{leakage, datapath}}$ & \SI{0.70}{\micro \watt} & \SI{0.70}{\micro \watt} \\
    \midrule
    $P_{\textrm{total}}$ & \SI{2.97}{\micro \watt} & \SI{14.9}{\micro \watt} \\
    \bottomrule
  \end{tabular}
  \label{tab:cwu_power}
\end{table}

\begin{table*}
  \centering
    \caption{Comparison of state-of-the-art smart wake-up units}
    \resizebox*{\linewidth}{!}{
    \begin{tabular}{lccccccc}
    \toprule
                          & Cho2019 \cite{Cho2019} \textsuperscript{2, 3}       & Giraldo2020 \cite{Giraldo2020} \textsuperscript{2, 3} & Wang2020 \cite{Wang2020a} \textsuperscript{2, 3}   & Rovere2018 \cite{Rovere2018} \textsuperscript{3} & \bf{Vega CWU}\footnotemark[1]{} \\
    \midrule
    Applications          & VAD                                                              & Keyword Spott.                                                     & Slope Matching                                                   & General Purpose                           & General Purpose                 \\
    Technology            & 180nm                                                            & 65nm                                                               & 180nm                                                            & 130nm                                     & 22nm                            \\
    Power Envelope        & \SI{14}{\micro\watt}                                             & \SI{2}{\micro\watt}                                                & \SI{17}{\nano\watt}                                              & \SI{2.2}{\micro\watt}                     & \SI{2.97}{\micro\watt}          \\
    Classification Scheme & NN                                                               & LSTM, GMM                                                          & Threshold, Slope                                                 & Threshold Sequence                        & HDC                             \\
    Area                  & \textasciitilde\SI{3.7}{\square\milli\meter} \textsuperscript{4} & \textasciitilde\SI{0.4}{\square\milli\meter} \textsuperscript{4}   & \textasciitilde\SI{1.8}{\square\milli\meter} \textsuperscript{4} & \SI{0.011}{\square\milli\meter}           & \SI{0.147}{\square\milli\meter} \\
  \end{tabular}
  }
    \begin{tablenotes}
      \item \footnotemark[1]{} Power consumption is reported for a compute intensive language classification algorithm and a typical always-on classification algorithm for EMG data.
      \item \footnotemark[2]{} Although these designs contain an application specific analog frontend (AFE) their digital NN accelerators could potentially be used for other applications in a smart wakeup scenario.
      \item \footnotemark[3]{} For fair comparison with our digital-only CWU, only the power consumption of the included general-purpose classification logic is considered.
      \item \footnotemark[4]{} Exact area breakdown is not available thus we estimated the area of the classification logic from the chip micrograph.
    \end{tablenotes}
  \label{tab:cwu-comparison}
\end{table*}

Table \ref{tab:cwu_power} summarizes the power consumption of the CWU within Vega; we measured the CWU's power operating at 32kHz when performing a real-time inference HDC algorithm on data received from 3 SPI peripherals (16 bit, 150 SPS/channel) and at 200 kHz (1 kSPS/channel).
At 32 kHz the CWU consumes \SI{2.97}{\micro\watt} of which 77\% is dynamic and 23\% leakage.
At 200 kHz, overall power consumption increases to \SI{14.9}{\micro\watt}.
It is worth noticing that the dynamic power consumption of the CWUs datapath (\SI{0.99}{\micro\watt} and \SI{6.21}{\micro\watt} respectively) is about 20\% lower than the dynamic power of CWU's SPI pads.

Table \ref{tab:cwu-comparison} provides an overview of recently published smart wake-up units.
Most of the state-of-the-art units implement application-specific wake-up algorithms such as KWS, VAD, EMG, although the designs mentioned in the table all contain classification circuitry that might be used for other applications \cite{Cho2019, Giraldo2020, Wang2020a}.
To the best of the author's knowledge, \textit{Hypnos} is, among entirely general-purpose solutions, the one providing the highest level of configurability thanks to its flexible digital implementation together with the versatility of HDC and the preprocessing chain, still featuring a similar power consumption with respect to the only other general-purpose solution implementing a less flexible threshold-sequence based wake-up \cite{Rovere2018}.

\subsection{Parallel Compute Cluster}

\begin{figure}[t]
  \centering
  \includegraphics[width=0.99\linewidth]{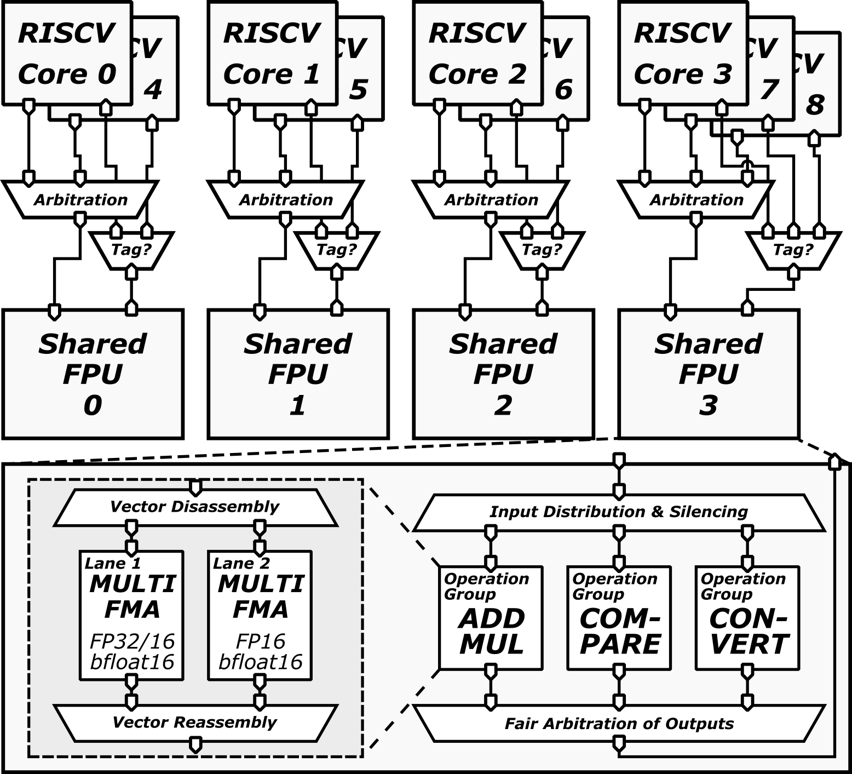}
  \caption{Architecture of the shared multi-precision FPU and its integration in the 9-cores cluster.}
  \label{fig:fpu}
  \vspace{-5mm}
\end{figure}

The programmable parallel accelerator of the system resides in a dedicated power and clock domain, communicating with the SoC subsystems with two AXI 4 ports (one master and one slave) connected to the SoC interconnect through dual-clock FIFOs. The cluster, built around 9 70kGE 4-pipeline stages RISCV cores, is turned on and adjusted to the required frequency when applications running on the FC offload computation-intensive kernels. The cores share data on a 128 kB shared multi-banked L1 memory, implemented with 16 8kB SRAM cuts, through a 1-cycle latency logarithmic interconnect \cite{LINT}. Similar to the L2 SoC interconnect, the L1 interconnect implements a word-level interleaving scheme to evenly distribute the requests, minimizing access contentions towards the SRAM banks. The cluster L1 memory can serve 16 parallel memory requests with less than 10\% contention rate even on data-intensive kernels, delivering up to 28.8GB/s at 450MHz. The program cache, implemented with latch-based SCM to improve energy efficiency over energy-expensive SRAM cuts for high-intensity activity, is hierarchical and includes 8 512 B private per-core plus 4 kB of 2-cycle latency shared cache to maximize efficiency with data-parallel code. In the common usage scenario. the ninth core manages DMA transfers leaving the computational workloads to the other 8 cores. It also includes a larger 1 kB L1 instruction cache. The refill path from L1 to L1.5 can be bypassed to avoid polluting the shared L1.5 cache. Fine-grain parallel thread dispatching is accelerated by a dedicated hardware event unit, which also manages clock gating of idle cores waiting for synchronization and enables resuming execution in 2 cycles. 

The RISCV cores used in Vega feature extensions (RVC32IMF-Xpulp) for NSAAs \cite{RI5CY}, such as hardware loops, post-incremented LD/ST, Single Instruction Multiple Data (SIMD) operations such as dot products operating on narrow 16- and 8-bit data types. The cores share 4 Floating-Point Units (FPUs) supporting FP32, FP16, and bfloat operations (a.k.a. SmallFloat extensions), as well as conversion operations among the different supported formats, including cast-and-pack instructions required to efficiently support packed SIMD vector operations \cite{FPU}. Some of the key operations present in many NSAA accumulating data in a higher-precision format to avoid loss of precision, such as Multiplication and Fused-Multiply-Add (FMA) can be performed as multi-format instruction, taking the product of two 16-bit operands and returning a 32-bit single-precision result. Division and square root operations are also supported in a stand-alone shared unit (DIV-SQRT). 

The FPUs are shared through a low-latency interconnect managing access contention in hardware, allowing to share one FPU among multiple cores in a fully transparent way from a software perspective. As opposed to \cite{WOLF_JSSC_2019}, in Vega we employ a partial interconnect with a static mapping of FPUs to cores, where units 0, 1, 2, 3 are shared among cores 0 \& 4, 1 \& 5, 2 \& 6, and 3 \& 7 \& 8 such that a core always accesses the same physical FPU instance, as shown in Figure \ref{fig:fpu}. This choice limits the flexibility in sharing the FPUs across processors but reduces the complexity of the interconnect towards the FPUs which are on the critical path of the cluster, guaranteeing high compute efficiency thanks to the single-cycle latency behavior of FP instructions, as demonstrated in section \ref{sec:benchmarking}.

\begin{figure}[t]
  \centering
  \includegraphics[width=0.99\linewidth]{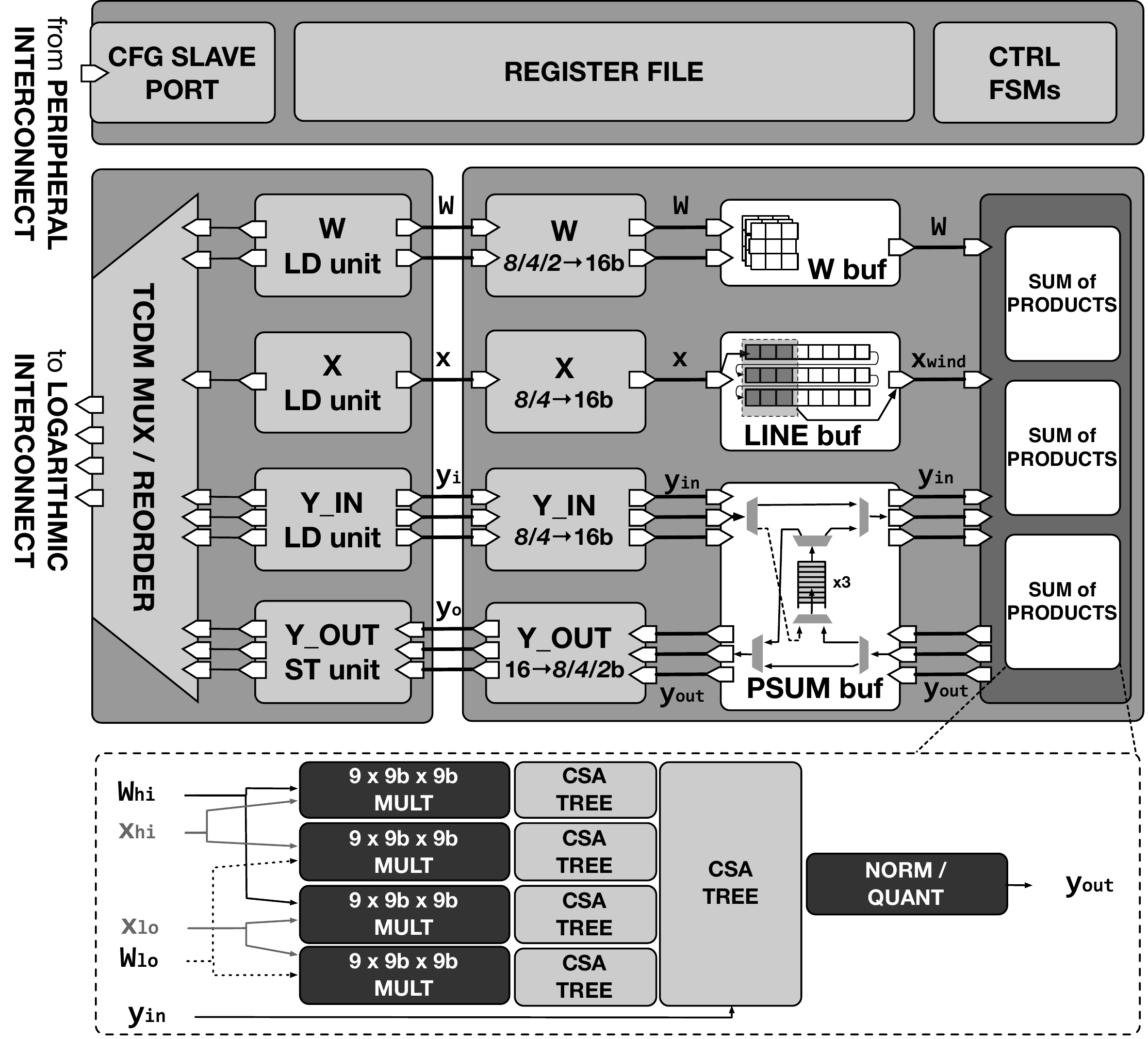}
  \caption{Hardware Convolution Engine (HWCE) microarchitecture.}
  \label{fig:hwce}
  \vspace{-5mm}
\end{figure}

CNN inference efficiency is boosted by a cluster-coupled machine learning hardware accelerator (HW Convolution Engine - HWCE) for multi-precision (4b/8b/16b) 3$\times$3 convolution, with 27 MACs in total. The microarchitecture of the HWCE is shown in Figure~\ref{fig:hwce}.
{The HWCE accesses the shared L1 TCDM memory by means of four 32-bit ports on the TCDM logarithmic interconnect. The HWCE load/store units, shown in Fig. 4 on the left side, convert between the TCDM memory protocol and a lightweight streaming protocol using ready/valid handshaking to manage stalls caused by memory contention. Bubbles in the data streams result in additional latency, but do not disrupt the functionality of the accelerator.}
The HWCE 
can be programmed via a set of memory-mapped registers via the peripheral interconnect.
A register shadowing mechanism enables to offload the next job without colliding with the currently running one.
In each accelerator job, the accelerator loads a set of up to three 3$\times$3 filters in an internal weight buffer; then, it starts streaming a continuous stream of input pixels from L1, which -- using an embedded line buffer -- are used to build a sliding window.

The stationary weights and the input sliding windows are upscaled to 16-bit and combined using three sum-of-product datapaths shared between all precision combinations. In each sum-of-products unit, 16-bits inputs and weights are first split into 9-bits sub-words (adding 1 bit for sign extension). The 9-bit sub-words are first combined along the 3$\times$3 spatial filter dimension, using four carry-save array (CSA) reduction trees. Then, they are combined using another CSA tree. In this way, even if the original inputs were smaller than the 16-bits upscaled version, fine-grain data and clock gating can be employed to disable leaves and branches of the reduction tree, minimizing activity and dynamic power.
Similarly, the datapath can also be reconfigured to implement 5$\times$5 convolutions by combining the three sum-of-products units and clock-gating unused units.
To further generalize to layers with partial or full input channel reuse, the accelerator includes a partial results FIFO buffer to accumulate convolution outputs from previous input channel contributions, either streamed-in from the L1 ports or from one of three internal partial sum buffers, acting as FIFOs. Symmetrically, the output of the three dot-product units can be either streamed out to L1 (possibly, after undergoing normalization and right-shift) or saved into the partial sum buffers to be re-used in the future.
Focusing on filter reuse, this design is particularly effective on popular VGG-style convolutional networks dominated by 3$\times$3 Conv layers~\cite{RepVGG} -- achieving up to 19 MAC/cycle on a 3$\times$3 convolutional layer.


%

\section{Chip Implementation and Measurements}

\begin{figure}[t!]
     \centering
     \includegraphics[width=\linewidth]{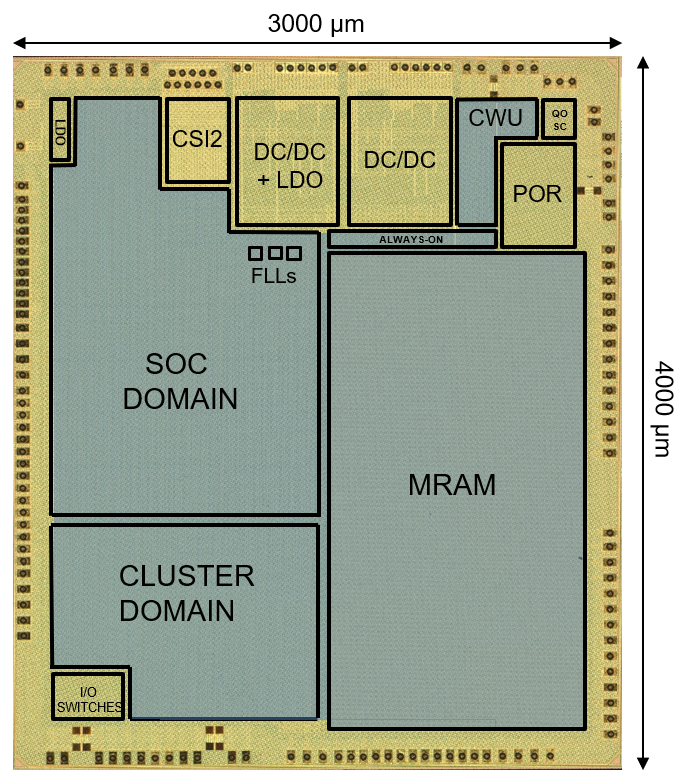}
     \caption{Vega SoC Die Micrograph.}
     \label{fig:chip}
\end{figure}

\begin{table}[t!]
\caption{Vega SoC Features.}
\centering
\begin{tabular}{cc}
    \hline
	Technology                                & CMOS 22nm FD-SOI     \\
	Chip Area                                 & 12mm$^2$             \\
	SRAM Memory                               & 1728 kB              \\
	MRAM Memory                               & 4 MB                 \\
	Equivalent Gates (NAND2)                  & 1.8 Mgates           \\
	Voltage Range                             & 0.6 V --  0.8 V      \\
	Frequency Range                           & 32 kHz -- 450 MHz    \\
	Power Range                               & 1.2 $\mu$W -- 49.4mW \\
	\hline
\end{tabular}
\label{tab:vega_features}
\end{table}

\begin{table}[t!]
\caption{Vega SoC Area Breakdown.}
\centering
\begin{tabular}{ccc}
    Instance           & Area [$mm^2$] & Percentage [\%] \\
    \hline
        MRAM           & 3.59 &	29.9 \\
        SoC Domain     & 2.69 &	22.4 \\
        Cluster Domain & 1.48 &	12.3 \\
        CWU            & 0.14 & 1.2 \\
        CSI2           & 0.15 & 1.2 \\
        DCDC1          & 0.36 & 3.0 \\
        DCDC2          & 0.36 & 3.0 \\
        POR	           & 0.14 & 1.1 \\
        QOSC           & 0.03 & 0.2 \\
        LDO            & 0.03 & 0.2 \\
	\hline
\end{tabular}
\label{tab:vega_area_breakdown}
\vspace{-5mm}
\end{table}

\begin{figure*}[t!]
     \centering
     \includegraphics[width=\linewidth]{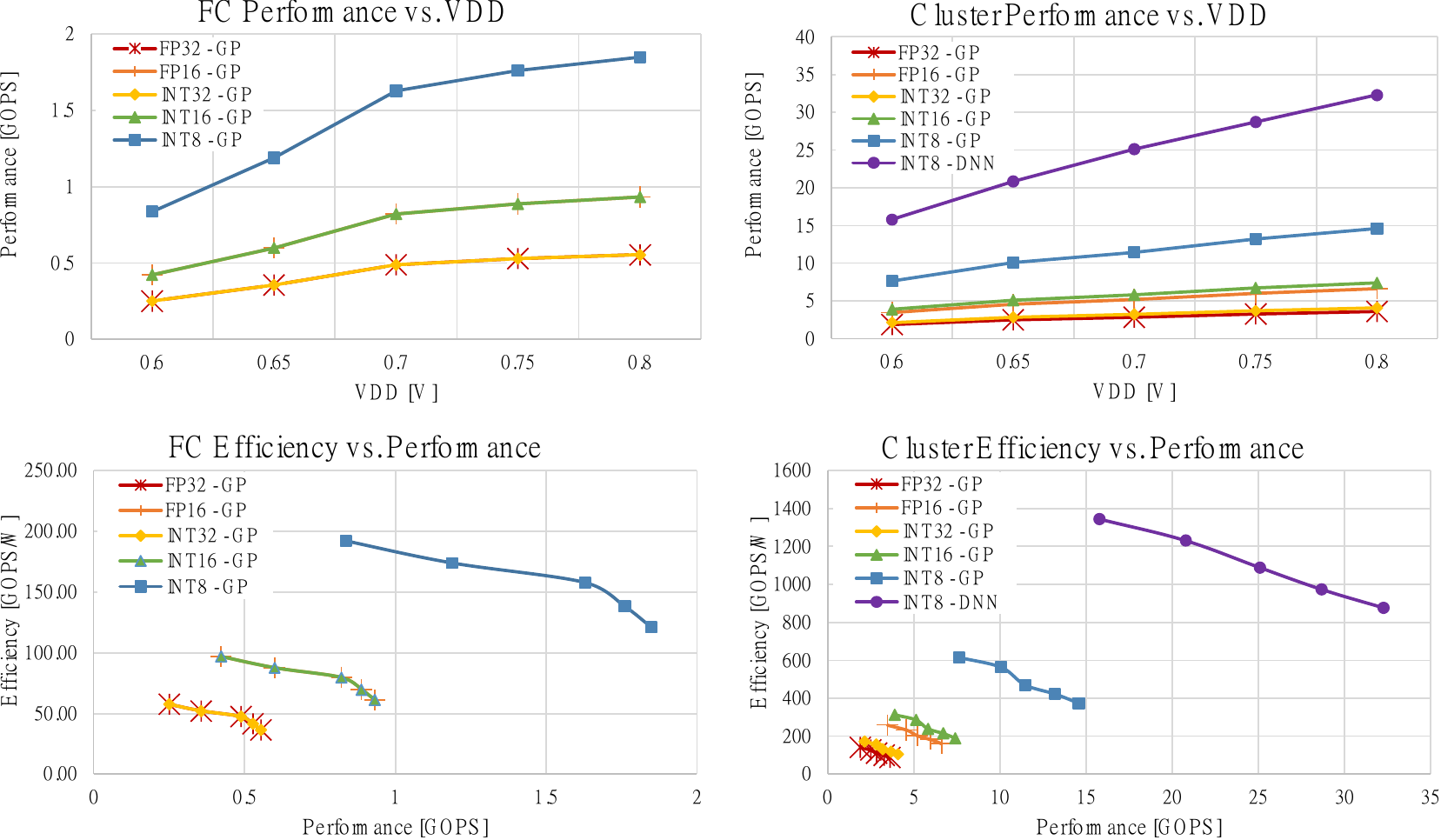}
     \caption{Vega SoC Performance and Efficiency.}
     \label{fig:perf}
     \vspace{-5mm}
\end{figure*}

Fig.~\ref{fig:chip} shows the microphotograph of the Vega SoC, along with its four main power domains described in Section \ref{sec:archi}. These power domains are managed by an additional always-on power domain operating from 0.6 V to 0.8 V, including commercial off-the-shelf components: two on-chip DC-DC converters, an LDOs operating at 3.6 V (VBAT), a power management unit used to switch on and off the other domains, a real-time clock (RTC) clocked by 1 MHz oscillator. The wake-up sources for the PMU are an external pad, the RTC, and the Cognitive Wake-up Unit (CWU). Three Frequency Locked Loops (FLLs) reside within the SoC domain and multiply the input clock generated by a kHz crystal oscillator (QOSC) to adjust the cluster domain and SoC domain frequency to the desired values. A third FLL generates the peripheral clock, which is then further divided and adjusted to serve the requirements of the different peripherals. Tab.~\ref{tab:vega_features} summarizes the main features of the Vega SoC, while Tab.~\ref{tab:vega_area_breakdown} summarizes the area breakdown of its main components. The largest blocks in the design are the 4 MB MRAM, the 1.6 MB of L2 memory within the SoC domain, and the power management IPs. On the other hand, the two programmable accelerators occupy less than 15\% of the overall SoC area.

\begin{figure}[t!]
     \centering
     \includegraphics[width=\linewidth]{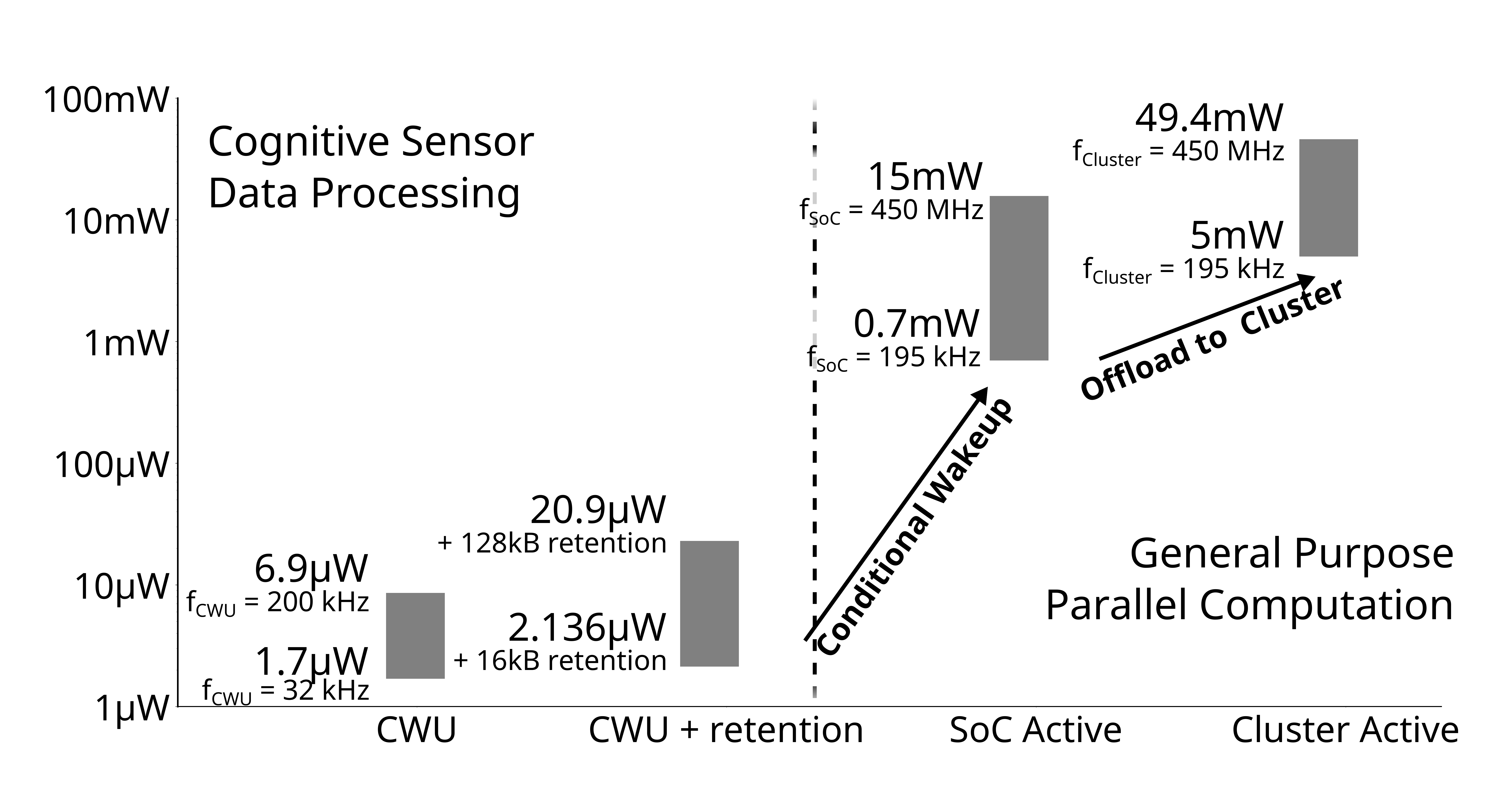}
     \caption{Vega SoC Power Modes and Consumption.}
     \label{fig:power_modes}
     \vspace{-7mm}
\end{figure}

Fig.~\ref{fig:power_modes} shows the power consumption of the components of the Vega SoC highlighted in Fig.~\ref{fig:chip} in the different power modes supported by the SoC, while Fig.~\ref{fig:perf} shows the matrix multiplication performance of the system in the active modes (FC active and cluster active) for all the supported data formats, from 8-bit integer to single-precision FP. In cognitive sleep mode, the CWU consumes from 1.7$\mu$W when operating at 32 kHz to 20.9$\mu$W when considering 128 kB of L2 memory to be turned into state-retentive mode. Once the SoC is turned on after a wake-up event, its power consumption goes from 0.7 to 15 mW, delivering up to 1.9 GOPS with an efficiency up to 200 GOPS/W  (8-bit integer). Finally, the SoC can offload highly compute-intensive tasks to the parallel cluster, delivering a much higher performance up to 15.6 GOPS with an efficiency of 614 GOPS/W on general-purpose processors (8-bit integer), and a performance of 32.2 GOPS with an efficiency up to 1.3 TOPS/W on convolutional workloads when HWCE is activated to accelerate the available software programmable processors, all within a power envelope of 49.4 mW. All the experiments were performed running on Vega an 8-bit matrix-multiplication kernel extracted from the PULP-NN library \cite{PULPNN} on the software-programmable cores and an 8-bit 3x3 convolution on the HWCE, assuming 2 operations for every MAC.

\section{Benchmarking}
\label{sec:benchmarking}

This section presents an extensive benchmarking of the Vega SoC, including both general-purpose NSAA exploiting 32-bit and 16-bit FP arithmetic as well as DNN workloads.

\subsection{Floating-Point NSAA Workloads}

\begin{table}[]
\caption{Benchmark suite. For each kernel we report its description and its FP intensity. Main application fields are listed in parentheses.}
\label{tab:fp_workloads_perf}
\resizebox{\columnwidth}{!}{\begin{tabular}{|c|c|c|}
\hline
\textbf{Application} &
\textbf{Description} &
\textbf{FP intensity} \\
\hline
MATMUL & Matrix multiplication (ExG, audio, image) & 57\%  \\
\hline
CONV   & Convolution kernel (ExG, audio, image)  & 55\%  \\
\hline
DWT    & Discrete Wavelet Transform (ExG) & 28\% \\
\hline
FFT    & Fast Fourier Transform (ExG, audio) & 63\% \\
\hline
FIR    & Finite impulse response filter (ExG) & 64\% \\
IIR    & Infinite impulse response filter (ExG) & 46\% \\
\hline
KMEANS & Unsupervised algorithm for data clustering (audio, image) & 83\% \\
\hline
SVM    & Supervised algorithm for data classification (audio, image)  & 35\% \\
\hline
\hline
\textit{\textbf{Average}} &    & \textit{\textbf{53\%}}  \\
\hline
\end{tabular}}
\end{table}

To assess the performance of diverse FP workloads, we considered a benchmark set including digital signal processing, machine learning algorithms, and basic linear algebra subroutines widely used in NSAA fields such as audio, image and ExG processing. To fully exploit the multi-precision support provided by the shared FPUs, we have extended the C/C++ frontend with two additional data types, namely \emph{float16} and \emph{bfloat16}. The compiler backend lowers scalar operations involving these additional types into the corresponding assembly instruction introduced by the smallFloat extensions (Section \ref{sec:archi}).
Packed-SIMD operations can be expressed through standard C/C++ operators on  GCC vector types. Vectorization of 16-bit data types reduces by a factor of two the execution time of the related FP operations. In addition, there is also a beneficial effect on memory bandwidth utilization since two 16-bit operands are read/written at the same time for each 32-bit memory access.

\begin{figure}[t!]
     \centering
     \includegraphics[width=\linewidth]{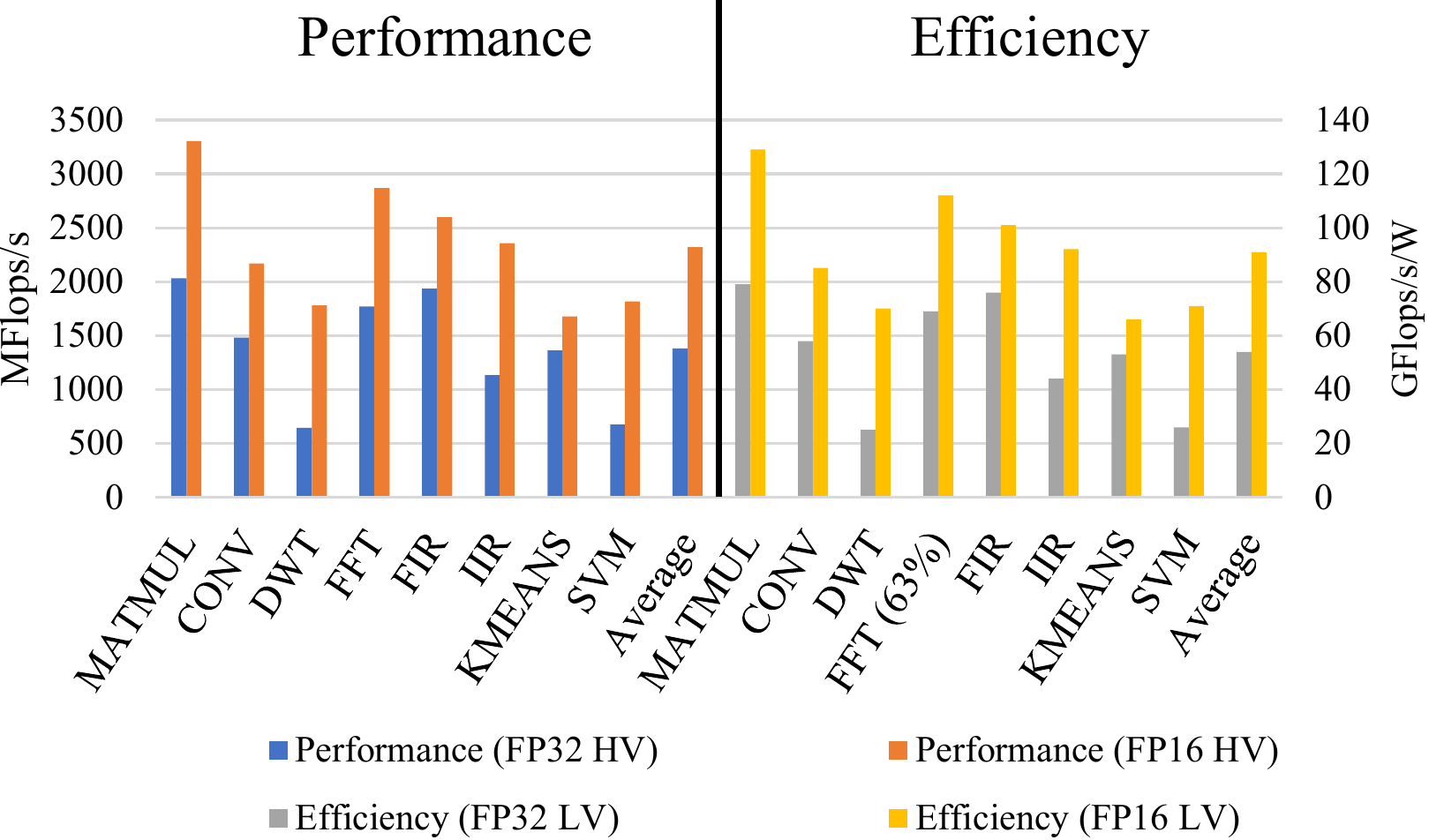}
     \caption{Performance and efficiency of FP NSAA.}
     \label{fig:perf_eff}
\end{figure}

Table~\ref{tab:fp_workloads_perf} reports description and FP intensity each kernel. FP intensity is the percentage of FP operations over the total number of instructions, computed at ISA level (i.e., on the kernel assembly code).
Fig.~\ref{fig:perf_eff} illustrates performance and efficiency of each kernel.
The performance metric reports how many millions of operations are completed per time unit considering FP32 and FP16 arithmetic for two operating points: 220 MHz, 0.6 V (LV) and 450 MHz, 0.8V (HV). These results demonstrate that the design choice of exploiting shared FPUs is not detrimental to the performance of FP workloads since programs include a mix of FP, ALU, control, and memory operations. The adoption of vectorial FP operations leads to an improvement of 1.46$\times$ over scalar ones. This value is lower than the theoretical speedup for the same effect discussed before: FP and memory operations take advantage of vectorization, while ALU and control operations are not affected. Some kernels (MATMUL, FFT, and FIR) are characterized by performance and efficiency gains higher than average values thanks to the use of fused multiply and accumulate instruction allowing to perform 2 FP operations per cycle. Thanks to the architecture design mapping integer and FP registers on a single register file, this gain is maintained for vectorial FP16 versions since programmers can manually optimize the code, including intrinsics for data packing and shuffling of vectors elements, with the final result to reduce the pressure on memory and shared FPUs.

\subsection{DNN NSAA Workloads and Data Flow}
\label{sec:DNN_benchmarking}


\begin{figure}[t]
     \centering
     \includegraphics[width=\linewidth]{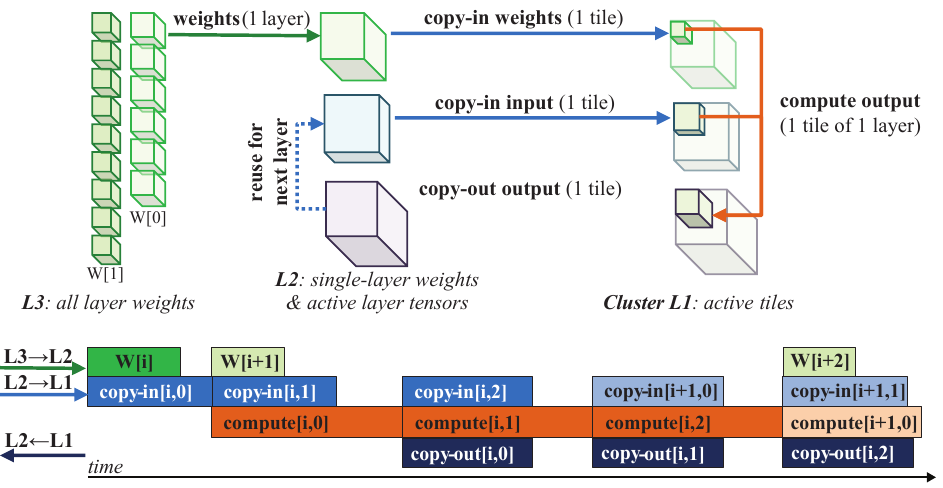}
     \caption{
     Example of DNN tiling software pipeline, showing the concurrent execution of weight data copies L3$\rightarrow$L2 (green); activation data copies L2$\leftrightarrow$L1 (blue); and compuation (orange) and the typical data flow in case of DNN tiling. Indeces distinguish layers $i$, $i+1$, etc. and tiles 0, 1, etc..}
     \label{fig:FIGURE5_tiling}
\end{figure}


\begin{figure}[t]
     \centering
     \includegraphics[width=\linewidth]{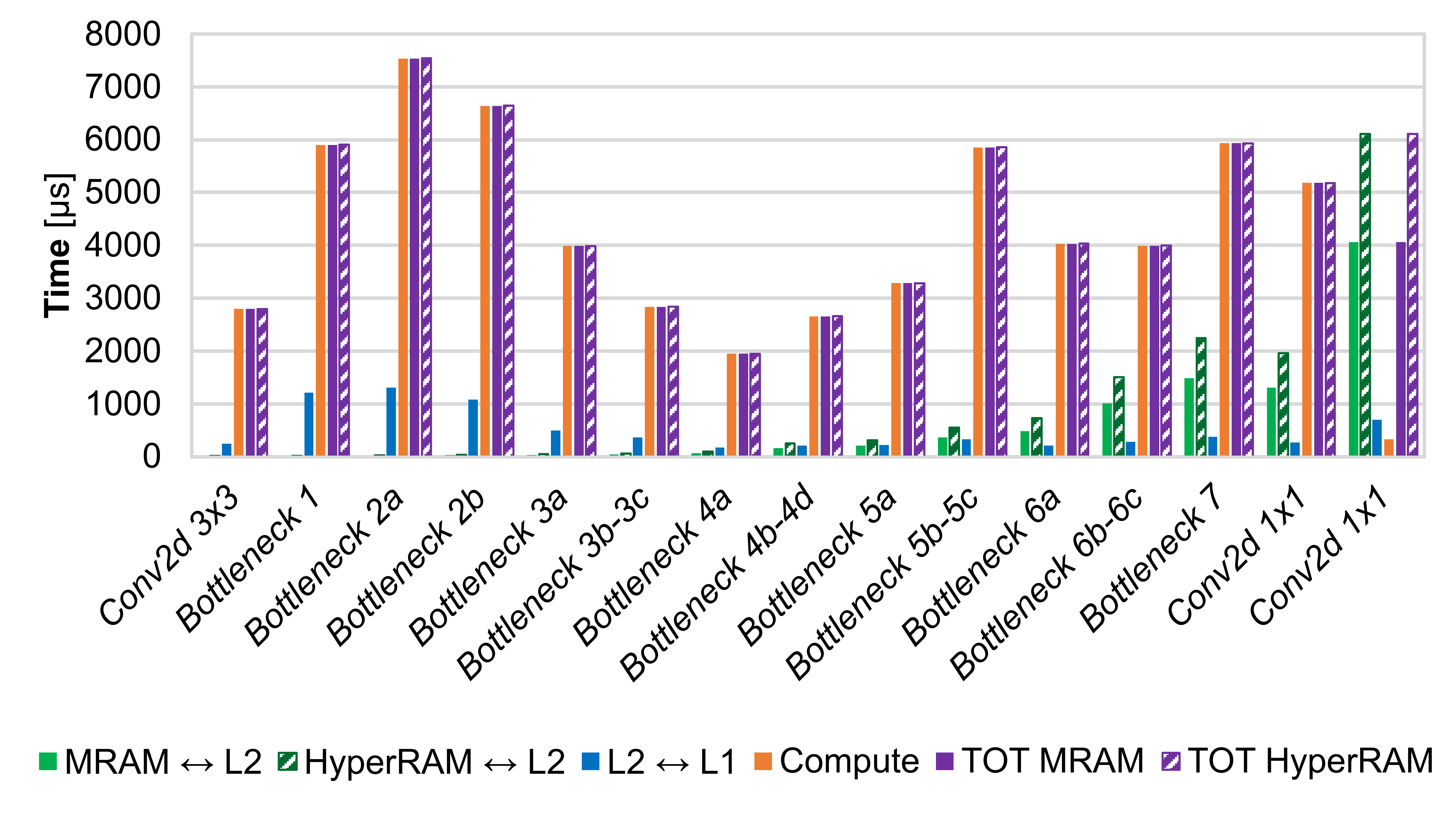}
     \caption{Layer-by-layer compute (orange), L2$\leftrightarrow$L1 (blue), and L3$\leftrightarrow$L2 (green) memory transfer latency in a \textit{MobileNetV2} use case. Total latency is shown in violet, highlighting MRAM (solid fill) and HyperRAM (dashed fill) as alternative choices. All layers except for the last one are compute-bound.
     Data reported in the nominal operating point with $f_{SOC}=\SI{250}{\mega\hertz}$, $f_{CL}=\SI{250}{\mega\hertz}$.}
     \label{fig:FIGURE5_time}
\end{figure}

\begin{figure}[t]
     \centering
     \includegraphics[width=\linewidth]{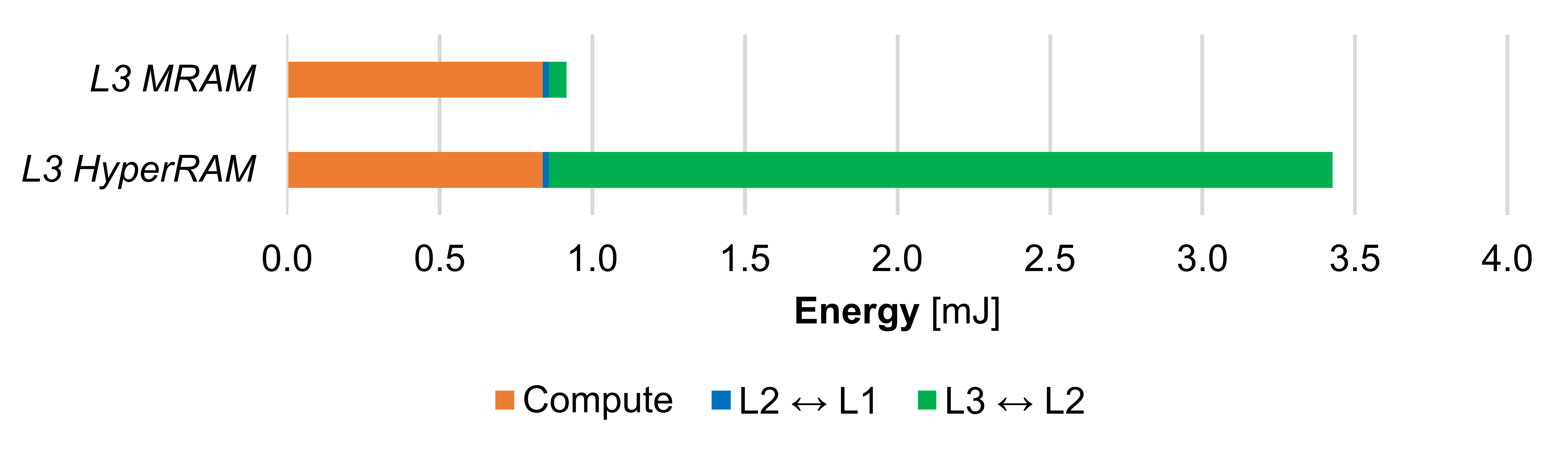}
     \caption{Comparison of total \textit{MobileNetV2} inference energy using weights allocated on external HyperRAM versus on-chip MRAM.}
     \label{fig:FIGURE5_energy}
\end{figure}

\begin{table}[t]
\caption{Comparison between the various data transfer channels used within a typical DNN execution in terms of available bandwidth and energy per byte.}
\begin{tabular}{rcc}
                                                    & \textit{Bandwidth [MB/s]} & \textit{Access Energy [pJ/B]} \\ \hline
\multicolumn{1}{l}{HyperRAM $\leftrightarrow$ L2} & 300                       & 20                            \\ \hline
\multicolumn{1}{l}{MRAM $\leftrightarrow$ L2}     & 200                       & 880                           \\ \hline
\multicolumn{1}{l}{L2 $\leftrightarrow$ L1}       & 1900                      & 1.4                           \\ \hline
\multicolumn{1}{l}{L1 access}                     & 8000                      & 0.9                           \\
\end{tabular}

     \label{fig:FIGURE5_bandwidth}
\vspace{-5mm}
\end{table}


The architecture of the Vega SoC is designed to be able to deploy realistically sized DNN for full on-chip inference, taking advantage of the MRAM for weight storage.
The general data flow for DNN inference is shown in Fig.~\ref{fig:FIGURE5_tiling}, focusing on a purely software execution pipeline based on the PULP-NN library, using 8-bit integers for all tensors (weights and activations).
Weights for all layers in the network are stored either in the on-chip MRAM (4 MB) or in an off-chip HyperRAM module -- both accessible via the I/O DMA, while intermediate activation tensors are allocated in the L2 shared memory (1.5 MB) and immediately deallocated after they are consumed by the following layer.
To enable computation on the cluster, both weights and input activation have to be divided into tiles that fit within the 128 KB of cluster L1 shared memory~\cite{dory}; 8 cores of the cluster are employed for actual computation, while the ninth is used as an \textit{orchestrator core} to manage data tiling and schedule data movement using the cluster DMA.
Computation is organized in a software pipeline with four stages, as shown in Fig.~\ref{fig:FIGURE5_tiling}:
\begin{enumerate}
    \item \textit{copy weights MRAM/HyperRAM$\rightarrow$L2}: weights for a full layer of the network are moved from MRAM/HyperRAM to L2 using the I/O DMA, programmed by the RISC-V FC core.
    \item \textit{copy-in input activations, weight L2$\rightarrow$L1}: the cluster orchestrator core schedules the copy of a weight and input tile from L2 to L1 using the cluster DMA.
    \item \textit{compute}: the 8 cluster compute cores, working entirely on 1-cycle latency L1, consume an input tile and a weight tile to produce a weight tile. We employ the PULP-NN library \cite{PULPNN} capable of achieving up to 15.5 MAC/cycle on 8 cores.
    3$\times$3 convolutional layers can alternatively employ the HWCE accelerator, achieving up to 27 MAC/cycle.
    \item \textit{copy-out output activations L1$\rightarrow$L2}: the cluster orchestrator core schedules the copy of the output tile from L1 to L2 using the cluster DMA.
\end{enumerate}
As shown in Fig.~\ref{fig:FIGURE5_tiling}, these four stages employ double-buffering and are fully overlapped so that throughput is dominated by the slowest stage. DORY~\cite{dory} is used both to calculate data tiling solutions fitting the memory constraints at all stages and to generate the orchestrator code.

To provide further insight into the data tiling scheme, we measured the bandwidth and the energy per byte for each of the data transfers described above. MRAM and HyperRAM results are measured on the silicon prototype, while L2/L1 accesses are estimated with power analysis using the final silicon netlist. The results are shown in Table~\ref{fig:FIGURE5_bandwidth}, in the nominal operating point ($Vdd_{SOC}=\SI{0.8}{\volt}$, $f_{SOC}=\SI{250}{\mega\hertz}$, $f_{CL}=\SI{250}{\mega\hertz}$).
MRAM and HyperRAM provide similar bandwidth, but thanks to on-chip integration, MRAM provides over 40$\times$ better energy efficiency. SRAMs (L2/L1) provide much higher bandwidth at a very low energy/byte but at a steep area cost and without state retention.

\begin{table*}[]
\caption{Vega performance and energy efficiency on RepVGG-A using SW or HWCE-based acceleration}
\label{tab:hwce_perf}
\centering
\begin{tabular}{l|c|cc|cc|c|c|c}
\multirow{2}{*}{}           & \textbf{ImageNet}            & \multicolumn{2}{c}{\textbf{Latency {[}ms{]}}} & \multicolumn{2}{c}{\textbf{Energy {[}mJ{]}}} & \multirow{2}{*}{\textbf{MMAC}} & \multirow{2}{*}{\textbf{Parameters {[}KB{]}}} & \multirow{2}{*}{\textbf{MRAM up to layer}} \\
                            & \textbf{Top-1 Acc. {[}\%{]}} & \textit{SW}        & \textit{HWCE} (speedup)  & \textit{SW} & \textit{HWCE} (eff. gain)      &                                &                                               &                                            \\ \hline
\textit{\textbf{RepVGG-A0}} & 72.41                        & 358                & 118 (3.03$\times$)       & 8.5         & 4.4   (+93\%)                  & 1389                           & 8116                                          & {stage 4, layer 12}                               \\ \hline
\textit{\textbf{RepVGG-A1}} & 74.46                        & 610                & 200 (3.05$\times$)       & 13.0        & 7.4   (+76\%)                  & 2364                           & 12484                                         & {stage 4, layer 6}                                \\ \hline
\textit{\textbf{RepVGG-A2}} & 76.48                        & 1320               & 433 (3.05$\times$)       & 25.7        & 15.8  (+63\%)                  & 5117                           & 24769                                         & {stage 4, layer 3}         \vspace{-1mm}    
\end{tabular}
\vspace{-5mm}
\end{table*}

As a complete case study for DNN inference on the Vega SoC, we selected \textit{MobileNetV2}~\cite{Sandler_2018_CVPR}: a widely used network topology for computer vision on mobile devices, used both for classification and object detection.
The MobileNetV2 template is very flexible, and it is also often employed as a template for tasks unrelated to vision~\cite{proxyless,mobilenet_app1,mobilenet_app2}.
The central block of MobileNetV2 is called a \textit{BottleNeck} and it consists of a sequence of three layers: a 1$\times$1 Convolutional \textit{expansion} layer, a 3$\times$3 Depthwise Convolution layer, and a 1$\times$1 Convolutional \textit{projection} layer. Additionally, the input of the expansion layer may be connected to the output of the projection layer by means of an additive residual connection. For our experiments, we employed the standard MobileNetV2 with depth multiplier 1.0 and input size 224$\times$224, which employes a total of 16 bottleneck layers with 7 different parameter combinations, plus 3 other layers at the front and back end of the network.
Fig.~\ref{fig:FIGURE5_time} reports the layer-wise execution time in microseconds when running on Vega with the data flow explained previously, without using the HWCE. Layers and BottleNecks in the front of the network tend to have more intensive transfers for activations (because their size is larger than the others), resulting in more L2$\leftrightarrow$L1 traffic. Conversely, in the back end of the network MRAM transfers, caused by larger weight sizes, are more relevant. Nevertheless, all layers except for the final one are compute-bound by a considerable margin.

In Fig.~\ref{fig:FIGURE5_energy}, we consider the full inference compute time and energy for MobileNetV2, comparing fully on-chip execution with weights on MRAM and the ``legacy'' flow using HyperRAM for weights. We observe that the time per inference is essentially the same, and it is compatible with real-time computation at more than 10 frames per second.
{The small difference of \SI{3}{\milli\second} is related exclusively to the final layer. This is because, as shown in Fig.~\ref{fig:FIGURE5_time}, all layers apart from the final $1\times 1$ convolution are compute-bound by a significant margin: the 50\% bandwidth improvement enabled by the MRAM, therefore, applies only to this layer.}
The substantial difference, however, is related to the much lower energy cost for memory access. Whereas in the legacy flow, HyperRAM accesses account for almost 25\% of the overall energy, the capability to store full-network weights on MRAM reduces this cost by a factor of 40$\times$, making it almost negligible compared to computing energy. As a consequence, the total energy per inference drops by 3.5$\times$ -- from \SI{4.16}{\milli\joule} to \SI{1.19}{\milli\joule}.

The HWCE engine included in Vega is not designed to operate efficiently on networks dominated by 1$\times$1 convolutions, such as MobileNet-V2, where the combined effect of parallel execution and RI5CY extensions deliver very high software throughput.
In that use case, employing the HWCE on 3$\times$3 depth-wise convolutions would improve their speed by a factor of $\sim$3$\times$, but lead to a modest $\sim$5\% speedup on the overall network.
On the other hand, VGG-style networks dominated by 3$\times$3 Conv layers are ubiquitous in real-world applications and are currently experiencing a resurgence of popularity in the DL community~\cite{RepVGG}.
On such networks, the HWCE can provide a substantial performance boost with respect to software-based execution, thanks to its exploitation of filter data reuse.

To better showcase this point, in Table~\ref{tab:hwce_perf} we present results in terms of energy and latency for three networks of the recently presented RepVGG-A0~\cite{RepVGG} family, which have been demonstrated to be highly competitive with the State-of-the-Art in terms of trainability, speed, and accuracy.
They are divided into 5 stages composed of 1, 2, 4, 14, and 1 layers, respectively -- all implemented as 3$\times$3 convolutions, plus a final fully connected layer.
All three networks are too big to fit entirely within the on-chip MRAM, so we revert to greedy allocation -- we keep early layer weights in MRAM until they fit {(as reported in the rightmost column of Table~\ref{tab:hwce_perf})}, to exploit its higher energy efficiency, and then we allocate back-end layers in HyperRAM.

The results are presented for both HWCE-based and SW-based computation, using the same PULP-NN layers of the MobileNetV2 case study in the latter case.
Almost all layers are compute-dominated, except for the final fully connected layer.
In such conditions, HWCE-based execution delivers a 3$\times$ speedup over SW-based; and a 60-90\% boost in system-level energy efficiency, depending on how much is the energy impact of HyperRAM traffic.

\begin{table*}[]
\caption{Comparison With State Of The Art}
\label{tab:soa}
     \centering
     \resizebox{\textwidth}{!}{
     \begin{tabular}{c|c|c|c|c|c|c}
                                                                                                                       & \textit{\begin{tabular}[c]{@{}c@{}}RISC-V VP\\ Schmidt et al.\\ ISSCC 2021 \cite{RISC_V_VP}\end{tabular}} & \textit{\begin{tabular}[c]{@{}c@{}}SleepRunner\\ Bol et al.\\ JSSC 2021 \cite{SLEEPRUNNER_JSSC_2021}\end{tabular}} & \textit{\begin{tabular}[c]{@{}c@{}}SamurAI\\ Miro-Panades et al.\\ VLSI 2020 \cite{SamurAI}\end{tabular}} & \textit{\begin{tabular}[c]{@{}c@{}}Mr. Wolf\\ Pullini et al.\\ JSSC 2019 \cite{WOLF_JSSC_2019}\end{tabular}} & \textit{\begin{tabular}[c]{@{}c@{}}GAP8\\ Flamand et al.\\ ASAP 2018 \cite{GAP8ASAP}\end{tabular}} & \textit{\textbf{\begin{tabular}[c]{@{}c@{}}Vega\\ (this work)\end{tabular}}}                          \\ \hline
\textit{Technology}                                                                                                    & FinFET 16nm                                                                                               & \begin{tabular}[c]{@{}c@{}}CMOS 28nm\\ FD-SOI\end{tabular}                                                         & \begin{tabular}[c]{@{}c@{}}CMOS 28nm\\ FD-SOI\end{tabular}                                                & CMOS 40nm                                                                                                    & CMOS 55nm                                                                                          & \begin{tabular}[c]{@{}c@{}}CMOS 22nm\\ FD-SOI\end{tabular}                                            \\ \hline
\textit{Die Area}                                                                                                      & 24 mm${}^{2}$                                                                                             & 0.68 mm${}^{2}$                                                                                                    & 4.5 mm${}^{2}$                                                                                            & 10 mm${}^{2}$                                                                                                & 10 mm${}^{2}$                                                                                      & 12 mm${}^{2}$                                                                                         \\ \hline
\textit{Type}                                                                                                          & Vector Processor                                                                                          & MCU                                                                                                                & Heter. MCU                                                                                                & Parallel MCU                                                                                                 & Parallel + Heter. MCU                                                                              & Parallel + Heter. MCU                                                                                 \\ \hline
\textit{Applications}                                                                                                  & DSP                                                                                                       & IoT GP                                                                                                             & IoT GP + NSAA +DNN                                                                                        & IoT GP + NSAA                                                                                                & IoT GP + NSAA + DNN                                                                                & IoT GP + NSAA + DNN                                                                                   \\ \hline
\textit{CPU/ISA}                                                                                                       & RV64GC                                                                                                    & \begin{tabular}[c]{@{}c@{}}CM0DS\\ Thumb-2 subset\end{tabular}                                                     & \begin{tabular}[c]{@{}c@{}}1x RI5CY\\ RVC32IMFXpulp\end{tabular}                                          & \begin{tabular}[c]{@{}c@{}}9x RI5CY\\ RVC32IMFXpulp\end{tabular}                                             & \begin{tabular}[c]{@{}c@{}}9x RI5CY\\ RVC32IMXpulp\end{tabular}                                    & \begin{tabular}[c]{@{}c@{}}10x RI5CY\\ RVC32IMFXpulp + SF\end{tabular}                                \\ \hline
\textit{\begin{tabular}[c]{@{}c@{}}Embedded SRAM\\ (State Retentive)\end{tabular}}                                     & 4.5 MB                                                                                                    & 64 kB s.r.                                                                                                         & \begin{tabular}[c]{@{}c@{}}464 kB\\ 40 kB s.r.\end{tabular}                                               & \begin{tabular}[c]{@{}c@{}}64 kB (L1)\\ 512 kB s.r. (L2)\end{tabular}                                        & \begin{tabular}[c]{@{}c@{}}64 kB (L1)\\ 512 kB (L2)\end{tabular}                                   & \begin{tabular}[c]{@{}c@{}}128 kB (L1)\\ 1600 kB s.r. (L2)\end{tabular}                               \\ \hline
\textit{Embedded NVM}                                                                                                  & -                                                                                                         & -                                                                                                                  & -                                                                                                         & -                                                                                                            & -                                                                                                  & 4 MB MRAM                                                                                             \\ \hline
\textit{Wake-up Sources}                                                                                               & -                                                                                                         & WiC                                                                                                                & WuR, RTC, Int, GPIO                                                                                       & GPIO, RTC                                                                                                    & GPIO, RTC                                                                                          & GPIO, RTC, Cognitive                                                                                  \\ \hline
\textit{\begin{tabular}[c]{@{}c@{}}Sleep Power\\ SRAM Ret. Slp.\\ Power\\ (St. Ret. SRAM)\end{tabular}}                & -                                                                                                         & \begin{tabular}[c]{@{}c@{}}5.4 $\mu$W\\ 9.4 $\mu$W\\ (64 kB s.r.)\end{tabular}                                     & \begin{tabular}[c]{@{}c@{}}-\\ 6.4 $\mu$W\\ (40 kB s.r.)\end{tabular}                                     & \begin{tabular}[c]{@{}c@{}}72 $\mu$W\\ 76.5 - 108 $\mu$W\\ (32 kB - 512 kB s.r.)\end{tabular}                & \begin{tabular}[c]{@{}c@{}}3.6 $\mu$W\\ 30 $\mu$W\\ (512 kB s.r.)\end{tabular}                     & \begin{tabular}[c]{@{}c@{}}1.7 $\mu$W (CWU)\\ 2.8 - 123.7 $\mu$W\\ (16 kB - 1.6 MB s.r.)\end{tabular} \\ \hline
\textit{\begin{tabular}[c]{@{}c@{}}INT Precision\\ FP Precision\end{tabular}}                                          & \begin{tabular}[c]{@{}c@{}}64\\ FP64, FP32, FP8\end{tabular}                                              & \begin{tabular}[c]{@{}c@{}}32\\ -\end{tabular}                                                                     & \begin{tabular}[c]{@{}c@{}}8, 16, 32\\ -\end{tabular}                                                     & \begin{tabular}[c]{@{}c@{}}8, 16, 32\\ FP32\end{tabular}                                                     & \begin{tabular}[c]{@{}c@{}}8, 16, 32\\ -\end{tabular}                                              & \begin{tabular}[c]{@{}c@{}}8, 16, 32\\ FP32, FP16, bfloat\end{tabular}                                \\ \hline
\textit{Supply Voltage}                                                                                                & 0.55 - 1V                                                                                                 & 0.4 - 0.8V                                                                                                         & 0.45 - 0.9V                                                                                               & 0.8 - 1.1V                                                                                                   & 1 - 1.2V                                                                                           & 0.5 - 0.8V                                                                                            \\ \hline
\textit{Max Frequency}                                                                                                 & 1.44 GHz                                                                                                  & 80 MHz                                                                                                             & 350 MHz                                                                                                   & 450 MHz                                                                                                      & 250 MHz                                                                                            & 450 MHz                                                                                               \\ \hline
\textit{Power Range}                                                                                                   & n.a. - 4 W                                                                                                & 5.4 - 320 $\mu$W                                                                                                   & 6.4 $\mu$W - 96 mW                                                                                        & 72 $\mu$@ - 153 mW                                                                                           & 3.6 $\mu$W - 75 mW                                                                                   & 1.7 $\mu$W - 49.4 mW                                                                                  \\ \hline
\textit{\begin{tabular}[c]{@{}c@{}}${}^{1,4}$Best Int Perf\\ ${}^{1,4}$Best Int Eff\\ ${}^{1,4}$@ Perf\end{tabular}}   & -                                                                                                         & \begin{tabular}[c]{@{}c@{}}31 MOPS (32b)\\ 97 MOPS/mW (32b)\\ @ 18.6 MOPS (32b)\end{tabular}                       & \begin{tabular}[c]{@{}c@{}}1.5 GOPS\\ 230 GOPS/W\\ @ 110 MOPS\end{tabular}                                & \begin{tabular}[c]{@{}c@{}}12.1 GOPS\\ 190 GOPS/W\\ @ 3.8 GOPS\end{tabular}                                  & \begin{tabular}[c]{@{}c@{}}6 GOPS\\ 79 GOPS/W\\ @ 3.5 GOPS\end{tabular}                            & \begin{tabular}[c]{@{}c@{}}15.6 GOPS\\ 614 GOPS/W\\ @ 7.6 GOPS\end{tabular}                           \\ \hline
\textit{\begin{tabular}[c]{@{}c@{}}${}^{2,4}$Best FP32 Perf\\ ${}^{2,4}$Best FP32 Eff\\ ${}^{2,4}$@ Perf\end{tabular}} & \begin{tabular}[c]{@{}c@{}}n.a\\ 92.3 GFLOPS/W\\ n.a\end{tabular}                                         & -                                                                                                                  & -                                                                                                         & \begin{tabular}[c]{@{}c@{}}1 GFLOPS\\ 18 GFLOPS/W\\ @ 350 MFLOPS\end{tabular}                                & -                                                                                                  & \begin{tabular}[c]{@{}c@{}}2 GFLOPS\\ 79 GFLOPS/W\\ @ 1 GFLOPS\end{tabular}                           \\ \hline
\textit{\begin{tabular}[c]{@{}c@{}}${}^{2,4}$Best FP16 Perf\\ ${}^{2,4}$Best FP16 Eff\\ ${}^{2,4}$@ Perf\end{tabular}} & \begin{tabular}[c]{@{}c@{}}368.4 GFLOPS\\ 209.5 GFLOPS/W\\ @ 73 GFLOPS\end{tabular}                       & -                                                                                                                  & -                                                                                                         & -                                                                                                            & -                                                                                                  & \begin{tabular}[c]{@{}c@{}}3.3 GFLOPS\\ 129 GFLOPS/W\\ @ 1.7 GFLOPS\end{tabular}                      \\ \hline
\textit{\begin{tabular}[c]{@{}c@{}}${}^{3,4}$Best ML Perf\\ ${}^{3,4}$Best ML Eff\\ ${}^{3,4}$@ Perf\end{tabular}}     & -                                                                                                         & -                                                                                                                  & \begin{tabular}[c]{@{}c@{}}36 GOPS\\ 1.3 TOPS/W\\ @ 2.8 GOPS\end{tabular}                                 & -                                                                                                            & \begin{tabular}[c]{@{}c@{}}12 GOPS\\ 200 GOPS/W\\ @ 7 GOPS\end{tabular}                            & \begin{tabular}[c]{@{}c@{}}32.2 GOPS\\ 1.3 TOPS/W\\ @ 15.6 GOPS\end{tabular}                         
\end{tabular}
\vspace{-1mm}
     }
    \begin{tablenotes}
      \item \footnotemark[1]{} 2 OPs 	 = 1 8-bit MAC on MatMul benchmark unless differently specified.
      \footnotemark[2]{} 2 FLOPSs = 1 FMAC on MatMul benchmark unless differently specified.
      \footnotemark[3]{} 8-bit ML Workloads.
      \footnotemark[4]{} Execution from SRAM.
    \end{tablenotes}
\end{table*}

\section{Comparison With S.o.A.}

Table \ref{tab:soa} shows a comparison with a wide range of programmable embedded computing platforms, including RISC-V based vector processors for transprecision FP computations \cite{RISC_V_VP} and low-power IoT computing systems \cite{SLEEPRUNNER_JSSC_2021} exploiting either parallelism \cite{WOLF_JSSC_2019}, heterogeneity \cite{SamurAI}, or both \cite{GAP8ASAP} to address the high computing requirements of emerging NSAA applications and DNNs.

The work presented in \cite{RISC_V_VP} proposes a RISC-V vector processor composed of 8 clusters, each one including a 64-bit scalar core coupled with a vector accelerator supporting double-, single-, and half-precision FP operations, with a maximum of 8, 16, and 32 operations per cycle, respectively. While the absolute performance of \cite{RISC_V_VP} is much higher than Vega, thanks to the significantly larger area and higher operating frequency, its peak energy efficiency on a matrix multiplication benchmark is only 1.62x and 1.16 better than the one of Vega for FP16 and FP32 operation, respectively, despite the more scaled technology node. Moreover, the efficiency of a vector processor is well known to significantly degrade when dealing with small datasets and irregular patterns typical of NSAA. In this work, we demonstrate leading-edge FP efficiency on a wide range of 32-bit and 16-bit FP NSAA thanks to the flexibility of the proposed software-programmable cluster, as shown in Section \ref{tab:fp_workloads_perf}.

With respect to traditional fully programmable IoT endnodes such as \cite{SLEEPRUNNER_JSSC_2021}, which is representative for a wide range of MCUs based on CortexM0 or similar low-cost processors, Vega delivers orders of magnitude better performance and efficiency in active mode, enabling the execution of complex NSAA not feasible on such tiny systems. However, it should be noted that the main focus of most of research work on IoT MCUs is on optimization of low-power states such as state-retentive SRAMs or minimization of deep-sleep power and wake-up time from deep sleep, addressed in Vega with commercial off the shelf IPs exploiting standard techniques not being a key contribution of this work.

With respect to the more closely related works, namely multi-core IoT end-nodes \cite{WOLF_JSSC_2019, GAP8ASAP}, the proposed SoC delivers more than 1.3× better peak performance and more than 3.2$\times$ better peak efficiency on non-DNN NSAA workloads. The main improvements for these workloads come, on top of the more scaled technology node, from the more optimized integration between the prefetch buffer of the the cores and the hierarchical instruction cache. This cuts the critical path of the design with no loss in functional performance, lowering the power consumption and improving energy efficiency. For what concerns FP support, Vega is much more flexible and efficient than Mr.Wolf, delivering 2$\times$ better peak performance 4.3x better peak efficiency on 32-bit single-precision FP workloads, thanks to the highly optimized, shared FP unit offering support for single-cycle NSAA operations, such as fused multiply and accumulate. Moreover, Vega offers more flexibility in terms of support for low-precision FP formats (i.e., 16-bit and bfloat) further gaining performance and efficiency. With respect to the most efficient hardware-accelerated IoT end-nodes \cite{SamurAI}, Vega achieves similar energy efficiency on DNN inference workloads at 5.5$\times$ better performance. On non-DNN, NSAA workloads, Vega achieves 10$\times$ and 2.5$\times$ higher performance and energy-efficiency despite the 2 platforms employing the same RI5CY core \cite{RI5CY}. This gain is achieved thanks to the architectural efficiency of the parallel computing cluster over sequential solutions, also demonstrated in Fig.~\ref{fig:perf}.

Finally, to the best of the authors' knowledge, the proposed SoC is the only IoT end-node featuring a configurable 1.7~$\mu$W cognitive wake-up unit capable of fully on-chip execution of state-of-the-art mobile DNNs, such as MobileNetV2 and RepVGG, from non-volatile memory support.

\section{Conclusion}

We presented Vega, an always-on IoT end-node SoC featuring a 1.7 $\mathrm{\mu}$W fully retentive cognitive wake-up unit coupled with a power/performance/precision scalable SoC. The proposed SoC can achieve up to 32.2 GOPS (@ 49.4 mW) peak performance on NSAAs, including mobile DNN inference, exploiting 1.6 MB of state-retentive SRAM, and 4 MB of non-volatile MRAM. To meet the performance and flexibility requirements of NSAAs, the SoC features 10 RISC-V cores: one core for SoC and IO management and a 9-cores cluster supporting multi-precision SIMD integer and FP computation. Two programmable ML accelerators boost energy efficiency in sleep and active state, respectively. The proposed SoC can deliver a peak performance of 32 GOPS with an efficiency up to 1.3TOPS/W. The proposed SoC is capable of fully on-chip execution of state-of-the-art mobile DNNs such as MobileNetV2 and RepVGG-A0 at 1.19 mJ/Inference and 4.4 mJ/Inference, respectively.

\bibliographystyle{IEEEtran.bst}

%

\begin{IEEEbiography}[{\includegraphics[width=1in,height=1.25in,clip,keepaspectratio]{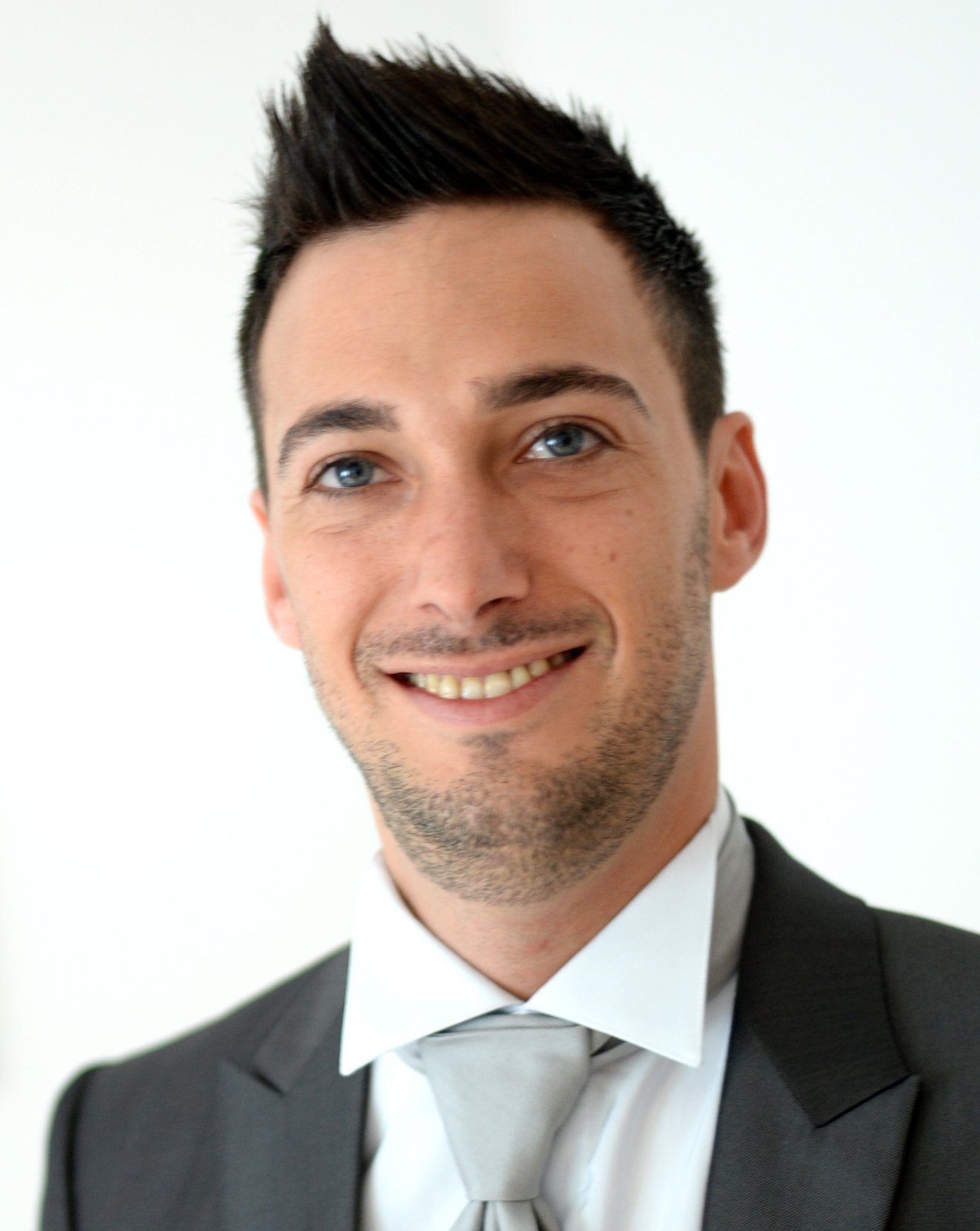}}]{Davide Rossi} received the Ph.D. degree from the University of Bologna, Bologna, Italy, in 2012. He has been a Post-Doctoral Researcher with the Department of Electrical, Electronic and Information Engineering “Guglielmo Marconi,” University of Bologna, since 2015, where he is currently an Assistant Professor. His research interests focus on energy-efficient digital architectures. In this field, he has published more than 100 papers in international peer-reviewed conferences and journals. He is recipient of Donald O. Pederson Best Paper Award 2018, 2020 IEEE TCAS Darlington Best Paper Award, 2020 IEEE TVLSI Prize Paper Award.
\end{IEEEbiography}

\begin{IEEEbiography}[{\includegraphics[width=1in,height=1.25in,clip,keepaspectratio]{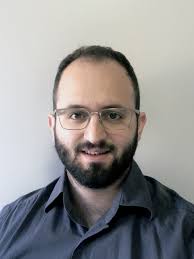}}]{Francesco Conti} received the Ph.D. degree in electronic engineering from the University of Bologna, Italy, in 2016. He is currently an Assistant Professor in the DEI Department of the University of Bologna. From 2016 to 2020, he held a research grant in the DEI department of University of Bologna and a position as postdoctoral researcher at the Integrated Systems Laboratory of ETH Zurich in the Digital Systems group. His research focuses on the development of deep learning based intelligence on top of ultra-low power, ultra-energy efficient programmable Systems-on-Chip. His research work has resulted in more than 40 publications in international conferences and journals and has been awarded several times, including the 2020 IEEE TCAS-I Darlington Best Paper Award.
\end{IEEEbiography}

\begin{IEEEbiography}[{\includegraphics[width=1in,height=1.25in,clip,keepaspectratio]{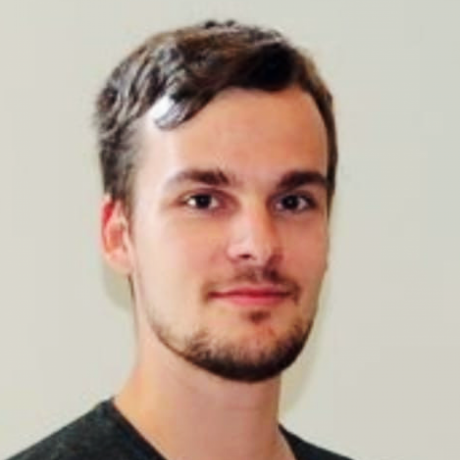}}]{Manuel Eggimann} Manuel Eggimann (Graduate Student Member, IEEE) received the B.Sc. and M.Sc. degrees in electrical engineering and information technology from ETH Zürich, Zürich, Switzerland, in 2018, where he is currently pursuing the Ph.D. degree with the ETH Zürich Integrated Systems Laboratory. His research interests include low-power hardware design, edge computing, and VLSI. Mr. Eggimann is a recipient of the best paper award at the 2019 IEEE 8th International Workshop on Advances in Sensors and Interfaces.
\end{IEEEbiography}

\begin{IEEEbiography}[{\includegraphics[width=1in,height=1.25in,clip,keepaspectratio]{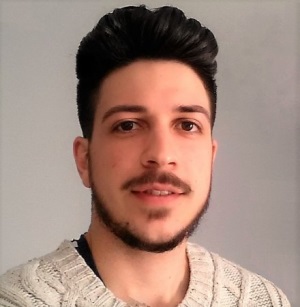}}]{Alfio Di Mauro} received the M.Sc.  degrees in Electronic Engineering from the Electronics and Telecommunications Department (DET) of Politecnico di Torino in 2016. Since September 2017, he is currently pursuing the Ph.D. at the Integrated System Laboratory (IIS) of the Swiss Federal Institute of Technology of Zurich. His research focuses on the design of digital Ultra-Low Power (ULP) System-on-Chip (SoC) for Event-Driven edge computing.
\end{IEEEbiography}

\begin{IEEEbiography}[{\includegraphics[width=1in,height=1.25in,clip,keepaspectratio]{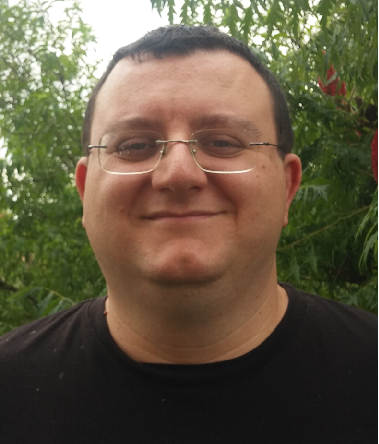}}]{Giuseppe Tagliavini} received a Ph.D. degree in Electronic Engineering from the University of Bologna, Bologna, Italy, in 2017. He is currently an Assistant Professor with the Department of Computer Science and Engineering (DISI) at the University of Bologna. He has co-authored over 40 papers in international conferences and journals. His main research interests include parallel programming models for embedded systems, run-time and compile-time optimization for multi/many-core platforms, HW/SW co-design of emerging computing architectures.
\end{IEEEbiography}

\begin{IEEEbiography}[{\includegraphics[width=1in,height=1.25in,clip,keepaspectratio]{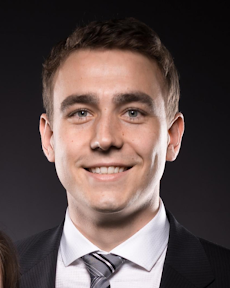}}]{Stefan Mach}
received his Ph.D. degree in Electrical Engineering from the Swiss Federal Institute of Technology Zurich (ETHZ), Switzerland, in 2021.
He is currently a Post-Doctoral Researcher with the Integrated Systems Laboratory at ETHZ. His research interests include transprecision computing, computer arithmetics, and energy-efficient processor architectures.
\end{IEEEbiography}

\begin{IEEEbiography} [{\includegraphics[width=1in,height=1.25in,clip,keepaspectratio]{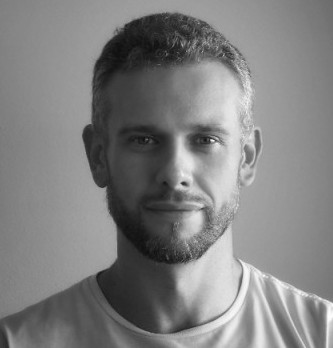}}]{Marco Guermandi} received his Ph.D. degree from the University of Bologna, Bologna, Italy, in 2009, and he has been carrying on research activity at the same University since then. He joined GreenWaves Technologies in 2019. His research interests focus on the development of energy efficient embedded systems, especially focusing on wearable devices for health monitoring and distributed sensor nodes. His research activity extends from IC design of analog IPs to the system level design of IoT devices, both from the hardware and software standpoint.
\end{IEEEbiography}

\begin{IEEEbiography} [{\includegraphics[width=1in,height=1.25in,clip,keepaspectratio]{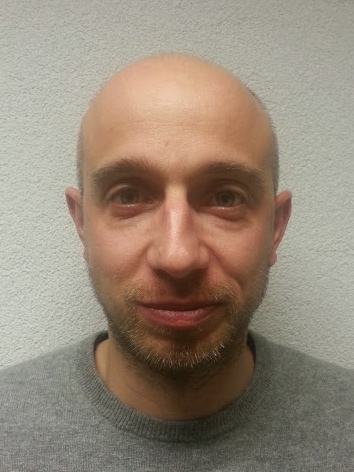}}]{Antonio Pullini} received the M.S. degree in electrical engineering from the University of Bologna, Bologna, Italy, and the Ph.D. degree from the Integrated Systems Laboratory, ETH Zürich, Zürich, Switzerland. He has been a Senior Design Engineer with iNoCs S.à.r.l., Lausanne, Switzerland, and he is currently with GreenWaves Technologies, Grenoble, France. His research interests include low-power digital design and networks on chip. In this field, he owns more than 50 papers in international peer-reviewed conferences and journals.
\end{IEEEbiography}

\begin{IEEEbiography} [{\includegraphics[width=1in,height=1.25in,clip,keepaspectratio]{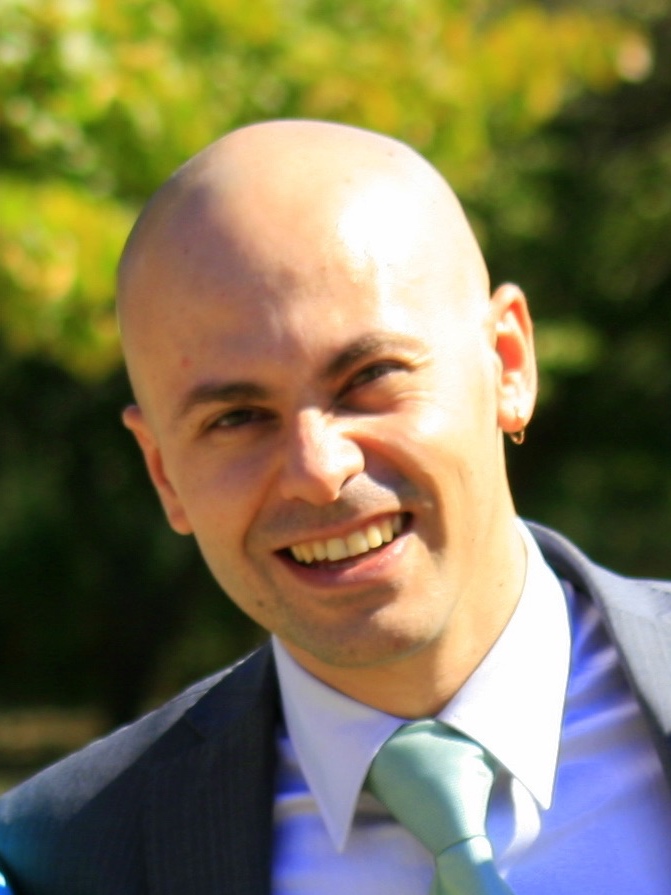}}]{Igor Loi} received the Ph.D. degree from the University of Bologna, Bologna, Italy, in 2010. He has been a Post-Doctoral Researcher with the
Department of Electrical, Electronic and Information Engineering ”Guglielmo Marconi,” University of Bologna, since 2006. He is currently with GreenWaves Technology, Crolles, France. His research activities are currently focused on ultralow power multi-core systems. In this field, he has published more than 40 papers in international peer-reviewed conferences and journals.
\end{IEEEbiography}

\begin{IEEEbiography} [{\includegraphics[width=1in,height=1.25in,clip,keepaspectratio]{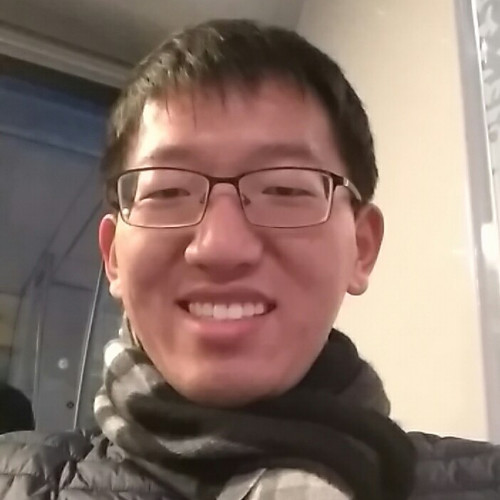}}]{Jie Chen} received MSc degree in electronic and information engineering from the Pierre and Marie Curie University, France, in 2017. He is now a PHD student in the Department of Electrical, Electronic and Information Engineering Guglielmo Marconi at the University of Bologna since 2018, at the same, he is working as ASIC designer for Greenwaves Technologies, France. His research activities are currently focused on ultra-low power multi-core systems, memory systems hierarchy and high speed memory interface.
\end{IEEEbiography}

\begin{IEEEbiography} [{\includegraphics[width=1in,height=1.25in,clip,keepaspectratio]{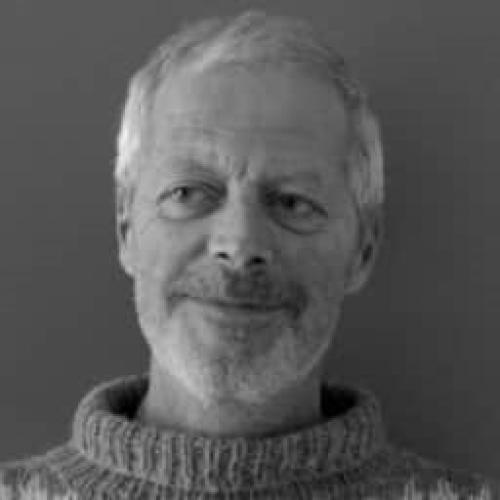}}]{Eric Flamand} Eric Flamand received the Ph.D. degree in computer science from INPG, Grenoble, France, in 1982. He was a Researcher with CNET, Grenoble, France, and CNRS, Grenoble, France, on architectural automatic synthesis, design and architecture, and compiler infrastructure for highly constrained heterogeneous small parallel processors. He then held different technical management positions within the semiconductor industry, first with Motorola, Chicago, IL, USA, where he was involved in the architecture definition and tooling of the StarCore DSP. Then with ST Microelectronics, Geneva, Switzerland, first being in charge of all the software development of the Nomadik Application Processor and then in charge of the P2012 corporate initiative aiming at the development of a many-core device. He is the co-founder and currently the CTO of Greenwaves Technologies, Villard-Bonnot, France, a French-based startup developing an IoT processor derived from the PULP project.
\end{IEEEbiography}

\begin{IEEEbiography}[{\includegraphics[width=1in,height=1.25in,clip,keepaspectratio]{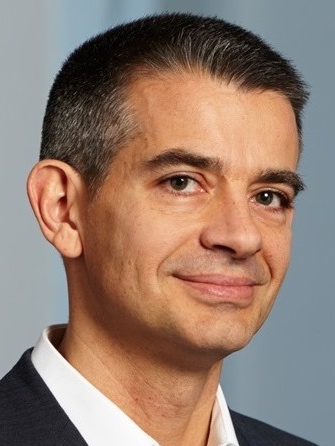}}]{Luca Benini}
received the Ph.D. degree in electrical engineering from Stanford University, Stanford, CA, USA, in 1997. He has served as the Chief Architect of the Platform 2012/STHORM Project with STMicroelectronics, Grenoble, France, from 2009 to 2013. He held visiting/consulting positions with École Polytechnique Fédérale de Lausanne, Stanford University, and IMEC. He is currently a Full Professor with the University of Bologna, Bologna, Italy. He has authored over 700 papers in peer-reviewed international journals and conferences, four books, and several book chapters. His current research interests include energy-efficient system design and multicore system-on-chip design. Dr. Benini is a member of Academia Europaea. He is currently the Chair of Digital Circuits and Systems with ETH Zürich, Zürich, Switzerland.
\end{IEEEbiography}








\end{document}